ORIGINAL ARTICLE

# A unified model of feed rotation in radio telescopes and GNSS antennas

**Joe Skeens[1] · Johnathan York[1] · Leonid Petrov[2] · Kyle Herrity[1] · Richard Ji-Cathriner[1] · Srinivas Bettadpur[3]**

**Abstract**

We describe a model that accounts for the phase rotation that occurs when a receiver or transmitter changes orientation while observing or emitting circularly polarized electromagnetic waves. This model extends work detailing Global Navigation Satellite Systems (GNSS) carrier phase wind-up to allow us to describe the interaction of changing satellite orientation with phase rotation in observing radio telescopes. This development is motivated by, and a critical requirement of, unifying GNSS and Very Long Baseline Interferometry (VLBI) measurements at the observation level. The model can be used for either stationary choke ring antennas or steerable radio telescopes observing either natural radio sources or satellites. Simulations and experimental data are used to validate the model and to illustrate its importance. In addition, we rigorously lay out the feed rotation correction for radio telescopes with beam waveguide and full Nasmyth focuses and validate the correction by observing the effect with dual polarization observations. Using this feed rotation model for beam waveguide telescopes, we produce the first phase delay solution for the VLBI baseline WARK30M–WARK12M. We provide a practical guide to using the feed rotation model in Appendix D.

## 1 Introduction

An electromagnetic wave in free space consists of orthogonal electric and magnetic fields oscillating perpendicular to each other and the direction of propagation. In linearly polarized light, the electric field oscillates along a fixed direction, while the magnetic field oscillates in a transverse plane. In circularly polarized light, the electric field can be described as the superposition of two perpendicular components with a quarter-cycle (90-degree) phase offset (Jackson (2021), Ch. 7). As a result, the electric field rotates in a circle at a rate proportional to the frequency, tracing a helical path as the light propagates. When tracking the phase of an incoming electromagnetic wave, a rotation in the orientation of a receiving (or transmitting) antenna along the plane perpendicular to the propagation direction causes an additional phase rotation and thus appears equivalent to an increase or decrease in the distance traveled during propagation (Thompson et al. (2001), Ch. 4). If an observer wishes to accurately measure distance using a precise phase measurement, the additional phase rotation due to this differential feed rotation must therefore be removed.

Our motivation to develop a differential feed rotation model for radio telescopes observing satellites comes from our work demonstrating interferometric observations of GNSS satellites and natural radio sources between a radio telescope and GNSS antennas on short baselines (Skeens

✉ Joe Skeens
jskeens1@utexas.edu

Johnathan York
york@arlut.utexas.edu

Leonid Petrov
leonid.petrov-1@nasa.gov

Kyle Herrity
kherrity@arlut.utexas.edu

Richard Ji-Cathriner
rcathriner@arlut.utexas.edu

Srinivas Bettadpur
srinivas@csr.utexas.edu

[1] Applied Research Laboratories, University of Texas at Austin, 10000 Burnet Rd, Austin, TX 78758, USA

[2] Goddard Space Flight Center, NASA, 8800 Greenbelt Rd, Greenbelt, MD 20771, USA

[3] Center for Space Research, University of Texas at Austin, 3925 W Braker Ln, Austin, TX 78759, USA

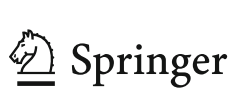

et al. 2023). The intent of the co-observation is to estimate a local tie vector directly between the reference points of the space geodetic instruments (Petrov et al. 2024). To accurately track phase when processing satellite observations with radio telescopes, a phase wind-up model accounting for the differential feed rotation between the observing radio telescope and the emitting satellite is necessary. In recent years, there has also been growing interest in geodetic VLBI processing of satellite observations by radio telescopes on long baselines as a method to realize frame ties between the GNSS and VLBI techniques (e.g., McCallum et al. (2024)). A third motivation is the planned European Space Agency satellite mission called GENESIS, which will carry an onboard VLBI transmitter and will serve as a collocation site for four space geodetic techniques—GNSS, VLBI, Satellite Laser Ranging (SLR), and Doppler Orbitography and Radiopositioning Integrated by Satellite (DORIS) (Delva et al. 2023). A simulation study has been published assessing the performance of VLBI observations of the proposed mission (Schunck et al. 2024).

In this manuscript, we will detail the phase wind-up correction for both satellite and natural radio source observations for four types of observing antennas:

- stationary GNSS antennas
- radio telescopes of all mount types with standard focuses (primary focus, Cassegrain, Gregorian)
- radio telescopes with full Nasmyth (FN) focuses
- radio telescopes with beam waveguide (BWG) focuses

Some modern radio telescopes use a complicated three-axis mount with dual linear polarization receivers, such as the Australian Square Kilometer Array Pathfinder (ASKAP) telescopes. The form of the correction for these mounts is outside the scope of this work.

The differential feed rotation effect is routinely accounted for in both the GNSS and VLBI communities but it is referred to by different names. The effect is commonly called carrier phase wind-up in the Global Navigation Satellite Systems (GNSS) community, and models must account for the rotation of both the transmitting satellite and the receiving antenna. GNSS satellites transmit right-handed circularly polarized (RHCP) signals in L band (1-2 GHz). The satellites continuously rotate to maintain their solar panels' orientation to the Sun. This orientation change combined with the movement of the satellite across the sky causes a large wind-up effect over the duration of a full pass for an observing antenna. In GNSS analysis, the phase wind-up correction must be computed for high-precision applications, for example Precise Point Positioning (PPP). Carrier phase wind-up was first described analytically in Wu et al. (1993), where the correction was derived by considering crossed dipole antennas for the receiver and transmitter and analyzing the voltage response in the receiving antenna assuming incident waves are perfectly RHCP. The expression was later improved by Beyerle (2009), who derived a second form of the correction that does not make the assumption of perfectly RHCP incident waves to track the phase rotation through ground reflections. The antennas typically used in these GNSS models are stationary, and the feed orientation is tracked in the local east–north–up frame.

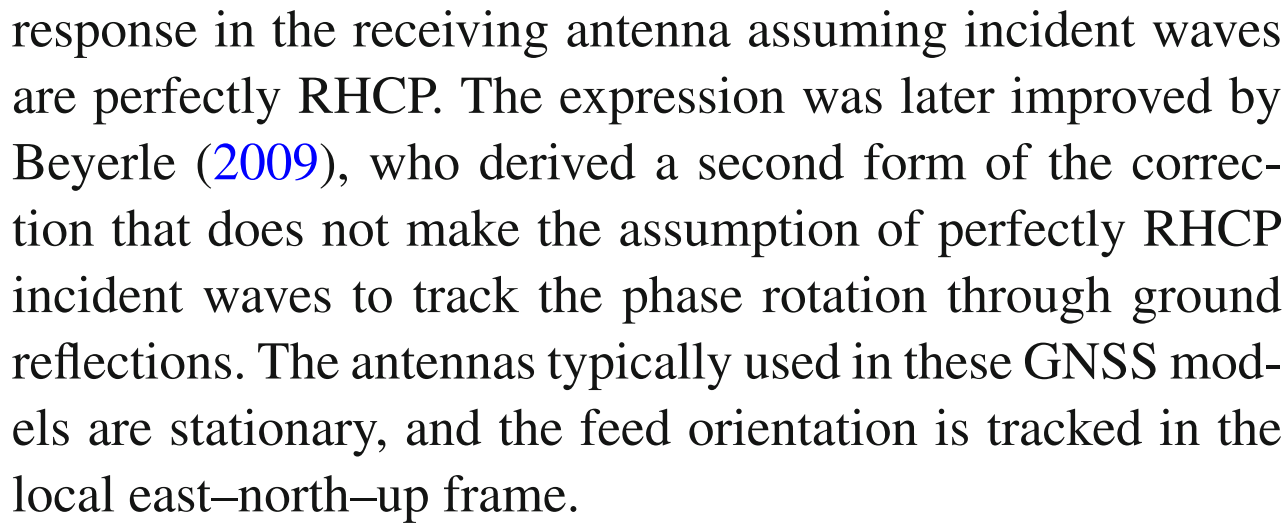

In contrast, the astronomy and Very Long Baseline Interferometry (VLBI) communities use large radio telescopes that slew to continuously track natural radio sources. These natural radio sources do not change their orientation on short timescales, so the observed phase rotation is caused only by the rotation of the feed in the observing radio telescope. The majority of radio telescopes have an azimuth–elevation mount type, meaning that the telescope slews by moving two independent axes that set the azimuth angle and elevation angle, respectively. In this most common case, the feed rotation correction is given by the parallactic angle (Cotton 1993); thus, the effect is frequently called the parallactic angle correction. However, the feed rotation correction we refer to here describes the rotation of incoming circularly polarized emission with respect to the feed horn, and as we will see, the parallactic angle correction is insufficient for azimuth–elevation mounted telescopes if the feed horn does not corotate with the telescope.

## 2 Theoretical foundations

In this manuscript, we are primarily concerned with the polarization state of electromagnetic radiation, which can equivalently be viewed in a variety of different polarization bases—most commonly either as a superposition of RHCP and left-handed circularly polarized (LHCP) emission, or as a superposition of horizontal and vertical linearly polarized emission. In this manuscript, we perform our analysis in the circular polarization basis, and we adopt the IEEE convention for RHCP, which specifies that the electric field rotates counterclockwise when viewed from the receiver and clockwise when viewed from the transmitter. For satellites emitting circularly polarized radio signals, the polarization purity is very high, meaning that the received emission is either almost all RHCP or all LHCP based on the designed polarization of the transmitter. Natural radio sources used as calibrators or for geodetic measurements are generally weakly polarized (Cotton 1993), meaning that there is a near even split between the power in the left and right hands, and to receive the maximum power from the source, both the RHCP and LHCP emission must be saved and processed.

The recorded phase of an antenna observing a radio source is determined by a number of terms. These include the distance between the observed source and the receiving antenna, $\phi_{\text{geom}}$, an offset caused by the bias of the clock used to timestamp the samples, $\phi_{\text{clock}}$, a phase delay due to the media the light has traveled through, $\phi_{\text{atm}}$, which typically includes a dispersive delay due to the ionosphere and a non-dispersive delay due to the neutral atmosphere, and finally a phase shift due to the relative orientation of the receiver and transmitter, $\phi_{\text{dfr}} = \Phi$, which we will refer to as phase wind-up or more generally as differential feed rotation.

$$\phi_{\text{obs}} = \phi_{\text{geom}} + \phi_{\text{clock}} + \phi_{\text{atm}} + \phi_{\text{dfr}} \tag{1}$$

In geodetic applications, the phase change due to the clock bias, $\phi_{\text{clock}}$, must be estimated, and these clock parameters absorb the absolute phase offset. For this reason, we only need to track the rotation of the receiver and transmitter with respect to a consistent direction in the plane of their feed and not the absolute offset from a particular reference direction unique to the hardware. However, it is critical to disambiguate the geometric phase, which is typically the desired measurement, from the additional phase rotation caused by the change in orientation of the receiver or transmitter.

To illustrate the origin of the phase correction due to wind-up, we will consider the simple case of a perfectly RHCP signal emitted by a transmitter and absorbed by a receiver oriented directly toward one another in parallel planes separated by a wavelength. The measured phase can be conceptualized as the angular distance of a reference direction in the plane of the receiver from the electric field vector projected to that plane. Because the direction of the electric field vector rotates as a function of distance, a change in the range by a fraction of a wavelength can be viewed as perfectly equivalent to a rotation in either the transmitter or receiver by the same fraction of a cycle. Figure 1 illustrates this ambiguity, where a measured phase of zero (upper left) can be disturbed to a quarter-cycle phase by either moving the receiver a quarter wavelength closer to the transmitter (upper right), rotating the receiver counterclockwise by a quarter cycle (lower left), or rotating the transmitter clockwise by a quarter cycle (lower right). When either the receiver or the transmitter changes orientation or moves transverse to the line of sight, the phase will change by both the expected geometric phase and an additional phase contribution from the projection of the electric field vector to the plane of the receiving antenna.

As with nearly all sources that deal with phase wind-up, we will provide models that deal with an idealized crossed dipole for the transmitter and receiver rather than including the effects of hardware biases and other complicated effects that usually can only be modeled empirically. A description of the interaction of phase wind-up with the antenna phase center calibration used in high-precision GNSS applications is shown in Rife et al. (2021), and a discussion of the calibration of instrumental effects in VLBI processing is available for example in Marti-Vidal et al. (2021). For LHCP emission, the direction of the helix drawn by the electric field is opposite, meaning that moving the receiver toward the transmitter results in a phase change of opposite sign.

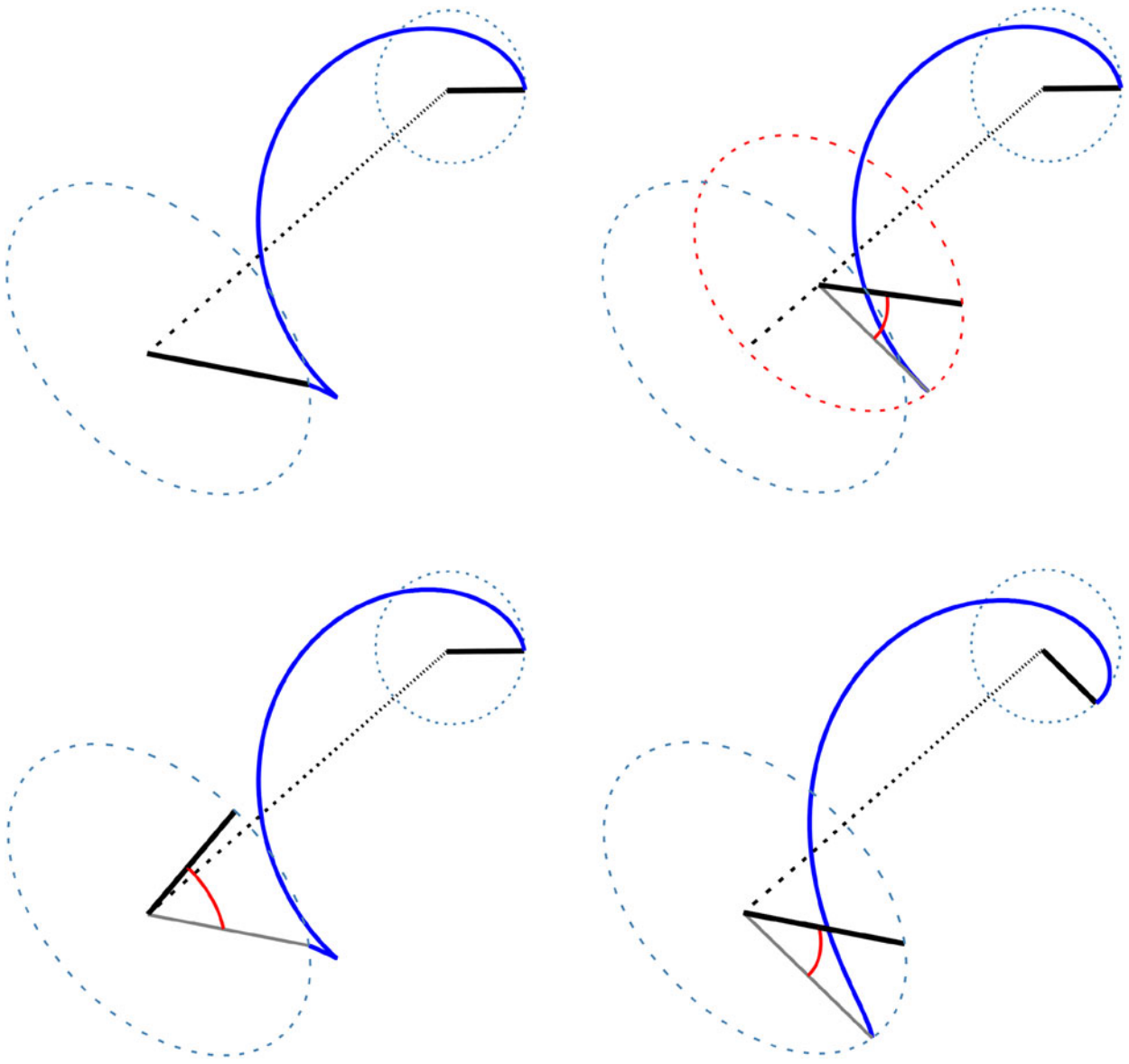

**Fig. 1** Relationship between the right-handed circularly polarized electric field vector (blue) as a function of distance and a reference direction on a transmitter and receiver that are in parallel planes and separated by one wavelength. Upper left: a reference scenario with a measured phase of zero. Upper right: the receiver is moved a quarter wavelength toward the transmitter. Lower left: the receiver is rotated counterclockwise by a quarter cycle. Lower right: the transmitter is rotated clockwise by a quarter cycle

## 2.1 The carrier phase wind-up model

To rigorously model the differential feed rotation or phase wind-up for a moving transmitter, we need an expression that can account for a receiver and transmitter with arbitrary relative orientations. As in Wu et al. (1993) and Beyerle (2009), we will track the rotation of the receiving and transmitting feeds by considering simple transverse, aligned, and boresight dipole vectors. These dipole vectors form the bases of a right-handed coordinate system for the receiver and transmitter. Figure 2 shows the definition of these vectors as well as the pointing vector from the satellite to receiver, $\hat{\boldsymbol{k}}$. The vector in the opposite direction from the receiver to the emitter is called the source unit vector and is more typically used in VLBI. The vector is defined as,

$$\hat{s} = -\hat{\boldsymbol{k}} = \frac{\boldsymbol{r}_{\text{sat}} - \boldsymbol{r}_{\text{ant}}}{\|\boldsymbol{r}_{\text{sat}} - \boldsymbol{r}_{\text{ant}}\|}, \tag{2}$$

where $\boldsymbol{r}_{\text{ant}}$ and $\boldsymbol{r}_{\text{sat}}$ are the Cartesian positions of the receiver and transmitter, respectively. In the context of the differential feed rotation, the effect of light-time correction on $\hat{s}$ is negligible, but the position of the satellite should in principle be at the time of emission.

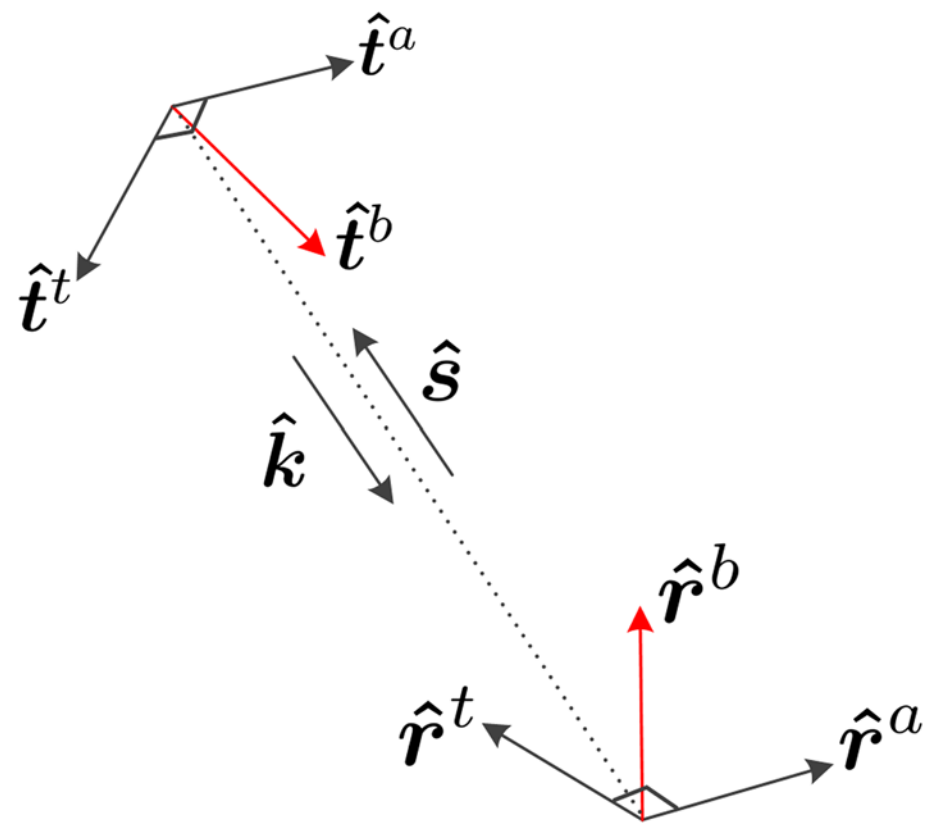

**Fig. 2** Transmitter and receiver aligned, transverse, and boresight dipole vectors as defined in Beyerle (2009)

The aligned and transverse dipole vectors of the receiver and transmitter as well as the source unit vector must be expressed in the same coordinate system. This coordinate system can be Earth-fixed, i.e., a realization of the International Terrestrial Reference System (ITRS) such as the International Terrestrial Reference Frame (ITRF), or it can be an inertial coordinate system such as the Geocentric Celestial Reference Frame (GCRF). For observations of GNSS satellites, the fixed system is generally easiest because receiver positions and orientations are obtained in an Earth-fixed system, and final high-fidelity satellite positions and orientations are distributed in an Earth-fixed system. In all simulations and data shown here, we used an Earth-fixed coordinate system to model the orientations of the receiver and transmitter.

From Beyerle (2009), the total phase wind-up due to the relative orientation of the transmitter and receiver feeds is given by,

$$
\begin{aligned}
\tan\Phi &= \frac{((\hat{\boldsymbol{k}} \times \hat{\boldsymbol{t}}^t) \times \hat{\boldsymbol{k}}) \cdot \hat{\boldsymbol{r}}^a + ((\hat{\boldsymbol{k}} \times \hat{\boldsymbol{t}}^a) \times \hat{\boldsymbol{k}}) \cdot \hat{\boldsymbol{r}}^t}{((\hat{\boldsymbol{k}} \times \hat{\boldsymbol{t}}^a) \times \hat{\boldsymbol{k}}) \cdot \hat{\boldsymbol{r}}^a - ((\hat{\boldsymbol{k}} \times \hat{\boldsymbol{t}}^t) \times \hat{\boldsymbol{k}}) \cdot \hat{\boldsymbol{r}}^t} \\
&= \frac{((\hat{\boldsymbol{s}} \times \hat{\boldsymbol{t}}^t) \times \hat{\boldsymbol{s}}) \cdot \hat{\boldsymbol{r}}^a + ((\hat{\boldsymbol{s}} \times \hat{\boldsymbol{t}}^a) \times \hat{\boldsymbol{s}}) \cdot \hat{\boldsymbol{r}}^t}{((\hat{\boldsymbol{s}} \times \hat{\boldsymbol{t}}^a) \times \hat{\boldsymbol{s}}) \cdot \hat{\boldsymbol{r}}^a - ((\hat{\boldsymbol{s}} \times \hat{\boldsymbol{t}}^t) \times \hat{\boldsymbol{s}}) \cdot \hat{\boldsymbol{r}}^t}.
\end{aligned} \tag{3}
$$

The phase wind-up $\Phi$ should be found through the sign-preserving four-quadrant arctangent function, atan2. Throughout this document, we will use the tangent and arctangent form for conciseness in the representation of the phase angles, while an implementation of the formulas should use the four-quadrant arctangent function.

For convenience, we define the operator that projects a vector to the plane perpendicular to the source unit vector $\hat{\boldsymbol{s}}$ as a matrix,

$$
\boldsymbol{P} = \boldsymbol{I} - \hat{\boldsymbol{s}}\hat{\boldsymbol{s}}^T. \tag{4}
$$

The phase wind-up can thus equivalently be expressed as,

$$
\tan\Phi = \frac{[\hat{\boldsymbol{r}}^a]^T \boldsymbol{P}\hat{\boldsymbol{t}}^t + [\hat{\boldsymbol{r}}^t]^T \boldsymbol{P}\hat{\boldsymbol{t}}^a}{[\hat{\boldsymbol{r}}^a]^T \boldsymbol{P}\hat{\boldsymbol{t}}^a - [\hat{\boldsymbol{r}}^t]^T \boldsymbol{P}\hat{\boldsymbol{t}}^t}. \tag{5}
$$

This projection matrix $\boldsymbol{P}$ has some useful properties: it is symmetric and idempotent (unchanged by a matrix multiplication with itself, $\boldsymbol{P}\boldsymbol{P} = \boldsymbol{P}$), and it can be expressed as the product of the skew-symmetric matrix $\boldsymbol{S}_\times$ and its transpose, $\boldsymbol{S}_\times \boldsymbol{S}_\times^T = \boldsymbol{P}$, where $\boldsymbol{S}_\times$ is the matrix equivalent of a cross product with $\hat{s}$:

$$
\boldsymbol{S}_\times = \begin{bmatrix} 0 & -s_z & s_y \\ s_z & 0 & -s_x \\ -s_y & s_x & 0 \end{bmatrix}. \tag{6}
$$

For an arbitrary vector $\boldsymbol{v}$, $\hat{\boldsymbol{s}} \times \boldsymbol{v} = \boldsymbol{S}_\times \boldsymbol{v}$. Note that Equations 3 and 5 from the Beyerle (2009) model do not necessarily represent a set of simple rotations about the line-of-sight vector. They represent the phase correction as observed by the canonical example of crossed dipole receiver and transmitter antennas. This model includes the effect of the responsiveness of an RHCP crossed dipole receiver to the LHCP component of an elliptically polarized signal that results from off-boresight emission by a crossed dipole RHCP transmitter. This sensitivity of the RHCP receiver to an LHCP signal would be called polarization leakage in the VLBI community, which is described with '$D$-terms' (see for example Marti-Vidal et al. (2021)).

If the receiver boresight vector, $\hat{\boldsymbol{r}}^b$, or the transmitter boresight vector, $\hat{\boldsymbol{t}}^b$, and the source unit vector $\hat{\boldsymbol{s}}$ are linearly dependent (coplanar), Equation 5 can be equivalently represented by the geometrically idealized phase wind-up model given in Wu et al. (1993). This equivalence is shown rigorously in Appendix A for $\hat{\boldsymbol{r}}^b = \hat{\boldsymbol{s}}$. In these cases, there is no longer a sensitivity to an LHCP component of the received signal, so the two models give identical phase wind-up corrections. The idealized phase wind-up model can be described as,

$$
\Phi = \operatorname{sgn}(\zeta)\operatorname{acos}\left(\frac{\boldsymbol{r} \cdot \boldsymbol{t}}{\|\boldsymbol{r}\|\|\boldsymbol{t}\|}\right) = \operatorname{atan2}(\zeta, \boldsymbol{r} \cdot \boldsymbol{t}), \tag{7}
$$

where

$$
\begin{aligned}
\boldsymbol{r} &= \boldsymbol{P}\hat{\boldsymbol{r}}^a - \boldsymbol{S}_\times \hat{\boldsymbol{r}}^t \\
\boldsymbol{t} &= \boldsymbol{P}\hat{\boldsymbol{t}}^a + \boldsymbol{S}_\times \hat{\boldsymbol{t}}^t \\
\zeta &= \hat{\boldsymbol{s}} \cdot (\boldsymbol{r} \times \boldsymbol{t}) = \boldsymbol{r}^T \boldsymbol{S}_\times^T \boldsymbol{t}
\end{aligned}. \tag{8}
$$

The vectors $\boldsymbol{r}$ and $\boldsymbol{t}$ are called effective dipole vectors for the receiver and transmitter, respectively. These vectors are by construction orthogonal to the source unit vector $\hat{\boldsymbol{s}}$. Tracking the relative orientation of these two vectors with a dot product is sufficient to track the phase wind-up. Because radio telescopes are steered such that $\hat{\boldsymbol{r}}^b = \hat{\boldsymbol{s}}$, the Wu et al. (1993) and Beyerle (2009) models are always equivalent for these systems.

For GNSS antennas observing GNSS satellites, Beyerle (2009) find the maximum phase error between the two models is on the order of a few mrad, equivalent to about 0.1 mm path error at L band frequencies. The error is small because the satellites are placed in Medium Earth Orbit and are steered to point at the Earth center, meaning $\hat{\boldsymbol{t}}^b \approx -\hat{\boldsymbol{s}}$. For an antenna on the surface of the Earth, the opening angle between the boresight vector and the line of sight does not exceed 13.9 degrees.

When accounting for phase wind-up in repeated (but not necessarily consecutive) observations of the same source, it is often necessary to preserve cycle continuity in the phase wind-up correction by adding the fractional cycle from Equation 3 or 7 to the nearest integer of the difference of the current and the previous wind-up value. For observation $i+1$ of source $j$,

$$\Phi_{i+1}^{j} = \Phi + 2\pi \left\lfloor \frac{\Phi_i^j - \Phi}{2\pi} \right\rceil. \tag{9}$$

This integer cycle correction is shown in Wu et al. (1993). If the observed source has changed orientation significantly since the previous observation, a more detailed phase unwrapping procedure may be necessary to preserve cycle continuity.

## 2.2 Feed rotation in radio telescope observations of natural radio sources

To track the receiver feed rotation for radio telescope observations of natural radio sources, we can use the phase wind-up model in Equations 7 and 8. We define $\boldsymbol{r}$ as a vector primitive that encodes the rotation of the observing feed horn. A convenient choice for this is the projection of the telescope's fixed-axis vector, $\hat{\boldsymbol{a}}$, to the receiver plane:

$$\boldsymbol{r}_{\text{tel}} = \boldsymbol{P}\hat{\boldsymbol{a}} = (\hat{\boldsymbol{s}} \times \hat{\boldsymbol{a}}) \times \hat{\boldsymbol{s}}. \tag{10}$$

It is critical that the source unit vector $\hat{\boldsymbol{s}}$ is in the same coordinate system as the receiver and transmitter effective dipole vectors. For natural radio source observations in existing analysis codes, it is likely that the source unit vector is expressed in inertial coordinates rather than Earth-fixed coordinates. In this case, $\hat{\boldsymbol{s}}$ must be transformed to the Earth-fixed coordinate system before calculating the feed rotation correction. Because we consider natural radio sources to be fixed in the inertial coordinate system, we must track the rotation of the receiver feed horn with respect to a vector fixed in the inertial frame. This fixed vector is chosen by convention to be the direction of the north celestial pole, $\hat{\boldsymbol{z}}$. To construct the effective transmitter dipole vector for natural radio sources, we also project $\hat{\boldsymbol{z}}$ to the plane normal to $\hat{\boldsymbol{s}}$:

$$\boldsymbol{t}_{\text{nat}} = \boldsymbol{P}\hat{\boldsymbol{z}} = (\hat{\boldsymbol{s}} \times \hat{\boldsymbol{z}}) \times \hat{\boldsymbol{s}}. \tag{11}$$

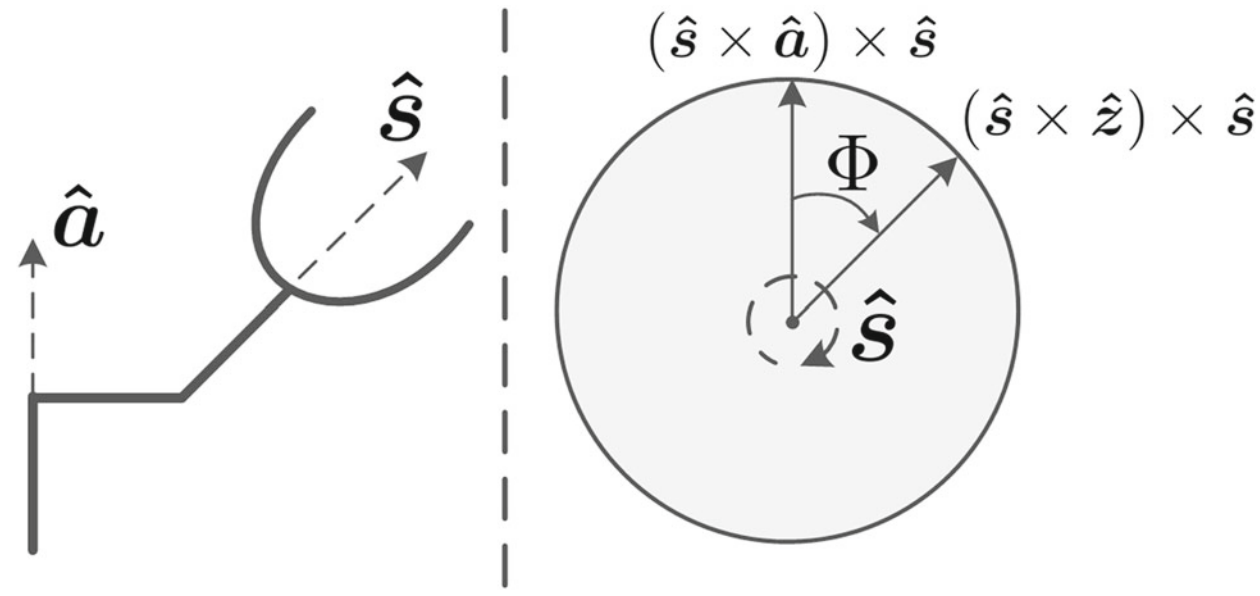

**Fig. 3** Left: A simple model of a radio telescope with fixed-axis vector $\hat{\boldsymbol{a}}$ observing a source with pointing vector $\hat{\boldsymbol{s}}$. Right: A view of the receiving paraboloid from the plane perpendicular to $\hat{\boldsymbol{s}}$. The phase correction for the receiver feed rotation is given by the angle $\Phi$

The receiver feed rotation is then given by the dot product of $\boldsymbol{r}$ and $\boldsymbol{t}$:

$$\begin{aligned} \Phi_{\text{tel, rx}} &= \operatorname{sgn}\left(\hat{\boldsymbol{s}} \cdot ((\boldsymbol{P}\hat{\boldsymbol{a}}) \times (\boldsymbol{P}\hat{\boldsymbol{z}}))\right) \operatorname{acos}\left(\frac{(\boldsymbol{P}\hat{\boldsymbol{a}}) \cdot (\boldsymbol{P}\hat{\boldsymbol{z}})}{\|\boldsymbol{P}\hat{\boldsymbol{a}}\|\|\boldsymbol{P}\hat{\boldsymbol{z}}\|}\right) \\ &= \operatorname{sgn}\left(\hat{\boldsymbol{a}}^T \boldsymbol{S}_\times^T \hat{\boldsymbol{z}}\right) \operatorname{acos}\left(\frac{\hat{\boldsymbol{a}}^T \boldsymbol{P}\hat{\boldsymbol{z}}}{\|\boldsymbol{P}\hat{\boldsymbol{a}}\|\|\boldsymbol{P}\hat{\boldsymbol{z}}\|}\right) \end{aligned} \tag{12}$$

Figure 3 shows the geometric origin of this equation. We can alternatively express the feed rotation using the tangent form to allow the use of the atan2 function to track both the sign and magnitude,

$$\Phi_{\text{tel, rx}} = \operatorname{atan} \frac{\hat{\boldsymbol{a}}^T \boldsymbol{S}_\times^T \hat{\boldsymbol{z}}}{\hat{\boldsymbol{a}}^T \boldsymbol{P}\hat{\boldsymbol{z}}}. \tag{13}$$

As Earth rotates, the relationship between the Earth-fixed vector, $\hat{\boldsymbol{a}}$, and the inertially fixed north celestial pole, $\hat{\boldsymbol{z}}$, changes. Additionally, when observing different radio sources, the projection of both of these vectors to the receiver plane changes. The feed rotation angle tracks the observed phase change due to both of these effects. Radio telescopes with different mount types have different fixed-axis vectors. Equation 13 accounts for the feed rotation in natural radio source observations of all mount types with conventional focuses, e.g., primary focus, Gregorian, Cassegrain. For RHCP observations, the feed rotation is subtracted from the phase measurements, and for LHCP observations, it is added.

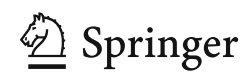

For the most common mount type called azimuth–elevation or altitude–azimuth, the fixed-axis vector is given by the local up vector,

$$\hat{\boldsymbol{a}}_{\text{az-el}} = \hat{\boldsymbol{u}}. \tag{14}$$

In this case, the phase wind-up is the parallactic angle, $\Psi$.

$$\Psi = \operatorname{sgn}\left(\hat{s} \cdot (\hat{\boldsymbol{u}} \times \hat{z})\right) \operatorname{acos}\left(\frac{(\boldsymbol{P}\hat{\boldsymbol{u}}) \cdot (\boldsymbol{P}\hat{z})}{\|\boldsymbol{P}\hat{\boldsymbol{u}}\| \|\boldsymbol{P}\hat{z}\|}\right) = \operatorname{atan}\frac{\hat{\boldsymbol{u}}^T \boldsymbol{S}_\times^T \hat{z}}{\hat{\boldsymbol{u}}^T \boldsymbol{P}\hat{z}}. \tag{15}$$

The parallactic angle, $\Psi$, is defined through spherical geometry as the angle that the great circle passing through the observed source and the local up direction makes with the great circle passing through the observed source and the celestial north pole. In terms of spherical angles, it is defined as (Cotton 1993),

$$\Psi = \operatorname{atan}\frac{\sin h}{\cos\delta \tan\phi - \sin\delta \cos h}. \tag{16}$$

where $h$ is the local hour angle of the observed source, $\delta$ is the topocentric declination of the source, and $\phi$ is the latitude of the observing telescope. Figure 4 shows the geometric correspondence between Equations 15 and 16. The parallactic angle is the most common correction applied to phase measurements, as the majority of radio telescopes have the azimuth–elevation mount type. For this reason, the feed rotation correction is commonly called the parallactic angle correction in the polarimetry literature despite the incorrectness of this label for other mount types.

For an equatorial/polar mount, the fixed-axis vector is aligned with the north celestial pole:

$$\hat{\boldsymbol{a}}_{\text{polar}} = \hat{z}. \tag{17}$$

From Equations 13 and 17, we can see that the feed rotation correction for the polar/equatorial mount type is 0, as $\hat{z}^T \boldsymbol{S}_\times^T \hat{z} = \hat{s} \cdot (\hat{z} \times \hat{z}) = 0$. The fixed-axis vector of radio telescopes with an X/Y (or XY) mount is in the direction of the X-axis. This is either the local north direction,

$$\hat{\boldsymbol{a}}_{\text{XY-N}} = \hat{\boldsymbol{n}}, \tag{18}$$

or the local east direction,

$$\hat{\boldsymbol{a}}_{\text{XY-E}} = \hat{\boldsymbol{e}}. \tag{19}$$

It is important to note that Equation 13 is not necessarily correct for radio telescopes with complicated internal reflections such as occur in beam waveguides. In Dodson (2009) and Dodson and Rioja (2022), the authors describe the feed rotation model for the full Nasmyth (FN) and beam waveguide (BWG) focus types. For these focus types, in contrast to the more popular primary focus, Cassegrain, and Gregorian focus types, additional terms are added by reflections off of mirrors that direct incoming light to a focus held fixed to a moving axis (FN) or to the antenna support structure (BWG). The only extant antennas of this type have azimuth–elevation mounts. The additional terms are therefore functions of the elevation and azimuth of the observing telescope.

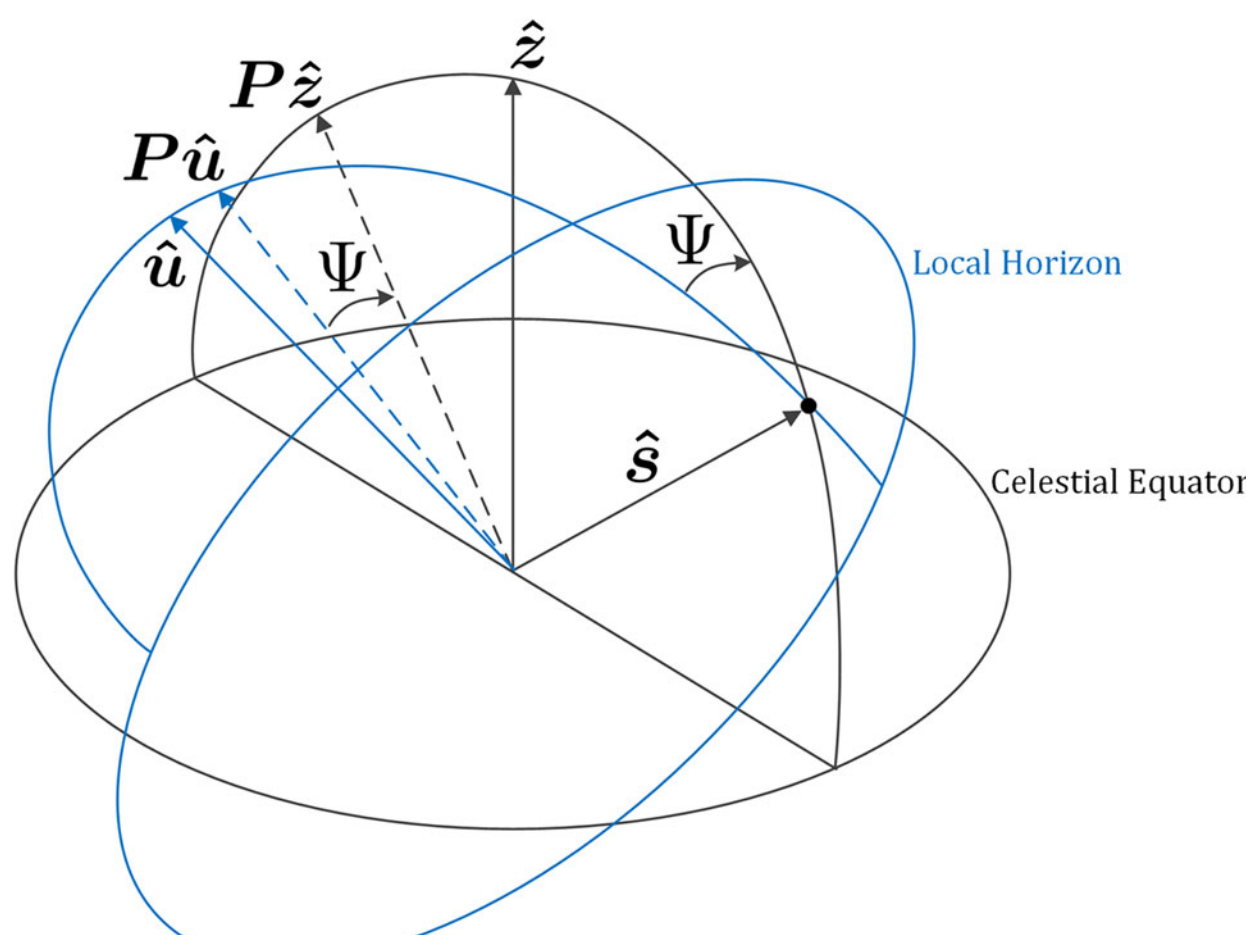

**Fig. 4** Geometric origin of the parallactic angle $\Psi$. The angle between the great circle connecting the local zenith and radio source with the great circle connecting the celestial north pole and the radio source is equivalent to the angle between the projections of the local zenith and the celestial north pole to the plane perpendicular to the source unit vector. As in Figure B.1 of Nothnagel (2024), the sign of the parallactic angle shown here is negative

There is little rigorous discussion of the feed rotation correction in the context of VLBI in the available literature. In Cotton (1993), the author outlines the full interferometric feed response for polarization-sensitive observations and briefly discusses the parallactic angle correction in this context. Chapter four of the seminal book Thompson et al. (2001) discusses the feed rotation correction in terms of the polarization ellipse of an observing antenna. Finally, Chapter 13 and Appendix B of Nothnagel (2024) show how to apply the parallactic angle correction in geodetic VLBI and provide a geometric interpretation.

## 3 Adapting the phase wind-up model for use with radio telescopes

Now we will consider observations of satellites with radio telescopes. In this case, both the receiver feed and transmitter feed rotate, leading to a differential feed rotation effect in the recorded phases. Starting from the formalism in Wu et al. (1993) (Equations 7 and 8), we show in Appendix B that the phase wind-up can be split into two terms, one accounting for the rotation and change in orientation of the transmitter antenna about the celestial north pole, $\Phi_{\mathrm{tx}}$, and the other accounting for the rotation and change in orientation of the receiver feed about the celestial north pole, $\Phi_{\mathrm{rx}}$:

$$\Phi = \Phi_{\mathrm{tx}} + \Phi_{\mathrm{rx}}. \tag{20}$$

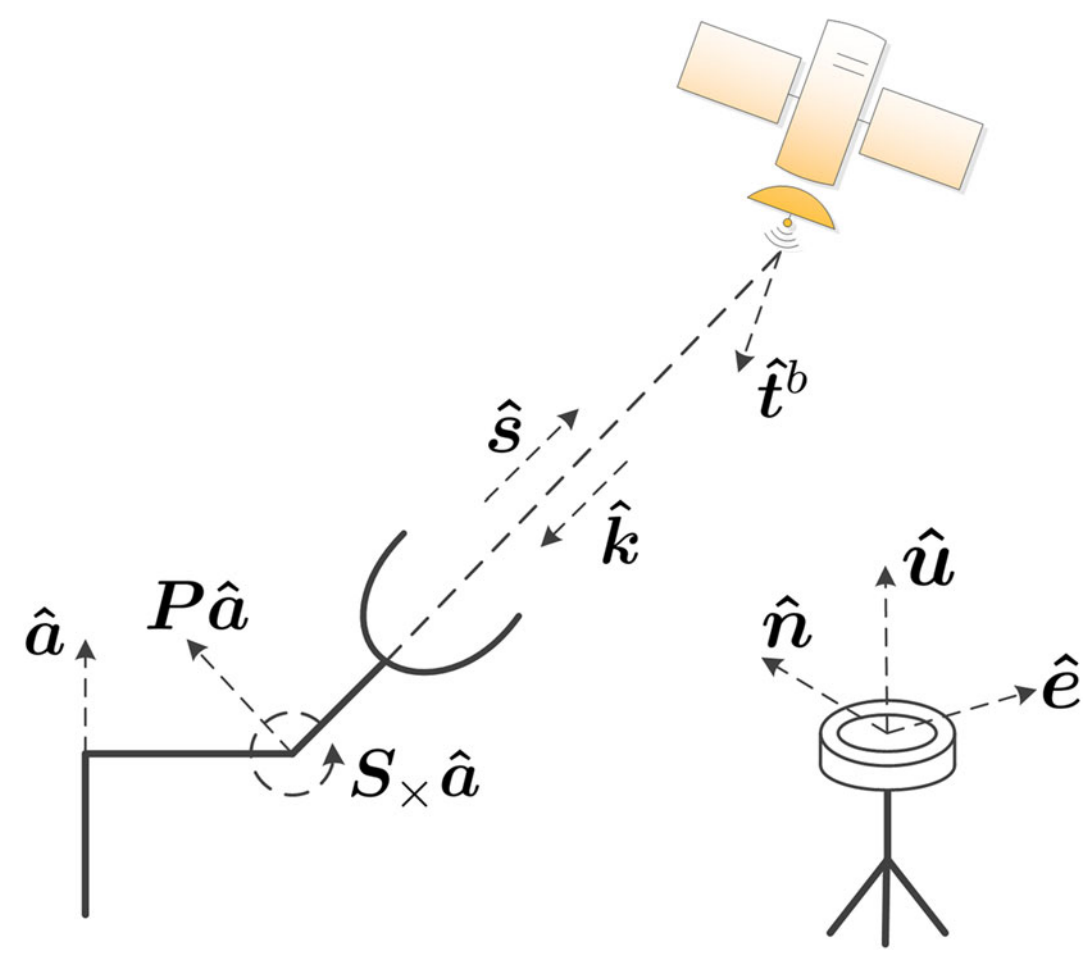

**Fig. 5** Orientation of a GNSS antenna and VLBI radio telescope observing a GNSS satellite

The transmitter rotation term is given by,

$$\tan \Phi_{\mathrm{tx}} = \frac{\hat{z}^T \boldsymbol{S}_\times^T \boldsymbol{t}}{\hat{z}^T \boldsymbol{t}}, \tag{21}$$

and the receiver rotation term is given by,

$$\tan \Phi_{\mathrm{rx}} = \frac{\boldsymbol{r}^T \boldsymbol{S}_\times^T \hat{z}}{\boldsymbol{r}^T \hat{z}}. \tag{22}$$

Note that when modeling an LHCP receiver observing an LHCP transmitter, the signs of $\Phi_{\mathrm{rx}}$ and $\Phi_{\mathrm{tx}}$ must be flipped.

### 3.1 Using the GNSS phase wind-up model for vertical dipole antennas and radio telescopes

While the application of phase wind-up corrections to observations with geodetic GNSS antennas has been well described, the phase wind-up (or differential feed rotation) of radio telescopes observing satellites has never been shown. To effectively compare the two antenna types, differential feed rotation corrections will be detailed for both receiving GNSS antennas and radio telescopes. Figure 5 shows a GNSS antenna pointing in the local up direction, $\hat{\boldsymbol{u}}$, and a radio telescope with a fixed-axis vector $\hat{\boldsymbol{a}}$ observing a GNSS satellite in the direction $\hat{s}$.

When observing GNSS satellites, the satellite attitude can be obtained through the high-rate orbit attitude files in ORBEX format supplied by the International GNSS Service. These files are estimated on a daily basis and given in the form of quaternions (Loyer et al. 2021). A quaternion consists of four values—one scalar and one vector triplet:

$$q_s(t) = q_s = \left(q_{s,0}\ \boldsymbol{v}_s\right) = \left(q_{s,0}\ q_{s,1}\ q_{s,2}\ q_{s,3}\right). \tag{23}$$

To apply these quaternion attitude files with the models detailed here, they can be transformed to the aligned and transverse dipole vectors through relatively simple functions of the quaternion elements:

$$\hat{\boldsymbol{t}}^a = \begin{bmatrix} q_{s,0}^2 + q_{s,1}^2 - q_{s,2}^2 - q_{s,3}^2 \\ 2(q_{s,1}q_{s,2} - q_{s,0}q_{s,3}) \\ 2(q_{s,1}q_{s,3} + q_{s,0}q_{s,2}) \end{bmatrix}, \tag{24}$$

$$\hat{\boldsymbol{t}}^t = \begin{bmatrix} 2(q_{s,1}q_{s,2} + q_{s,0}q_{s,3}) \\ q_{s,0}^2 - q_{s,1}^2 + q_{s,2}^2 - q_{s,3}^2 \\ 2(q_{s,2}q_{s,3} - q_{s,0}q_{s,1}) \end{bmatrix}. \tag{25}$$

In the absence of a high-fidelity quaternion attitude model, a simple model for a zenith-pointing satellite with Sun-oriented solar panels is given in Montenbruck et al. (2015), from which the boresight vector, transverse dipole vector, and aligned dipole vector can be found as,

$$\begin{aligned} \hat{\boldsymbol{t}}^b &= \frac{-\boldsymbol{r}_{\mathrm{sat}}}{\|\boldsymbol{r}_{\mathrm{sat}}\|} \\ \hat{\boldsymbol{t}}^t &= \frac{\hat{\boldsymbol{t}}^b \times \hat{\boldsymbol{e}}_\odot}{\|\hat{\boldsymbol{t}}^b \times \hat{\boldsymbol{e}}_\odot\|} \\ \hat{\boldsymbol{t}}^a &= \hat{\boldsymbol{t}}^t \times \hat{\boldsymbol{t}}^b = \frac{(\hat{\boldsymbol{t}}^b \times \hat{\boldsymbol{e}}_\odot) \times \hat{\boldsymbol{t}}^b}{\|\hat{\boldsymbol{t}}^b \times \hat{\boldsymbol{e}}_\odot\|} \end{aligned}. \tag{26}$$

The Sun pointing vector, $\hat{\boldsymbol{e}}_\odot$, is given by,

$$\hat{\boldsymbol{e}}_\odot = \frac{\boldsymbol{r}_{\mathrm{sun}} - \boldsymbol{r}_{\mathrm{sat}}}{\|\boldsymbol{r}_{\mathrm{sun}} - \boldsymbol{r}_{\mathrm{sat}}\|}. \tag{27}$$

The effective dipole vector $\boldsymbol{t}$ can be calculated from $\hat{\boldsymbol{t}}^a$ and $\hat{\boldsymbol{t}}^b$ using Equation 8.

For a GNSS antenna, the boresight vector is the local up vector. The local east and north vectors are the aligned and transverse dipole vectors for the receiving antenna:

$$\begin{aligned} \hat{\boldsymbol{r}}^b_{\mathrm{GNSS}} &= \hat{\boldsymbol{u}} = \frac{\boldsymbol{r}_{\mathrm{ant}}}{\|\boldsymbol{r}_{\mathrm{ant}}\|} \\ \hat{\boldsymbol{r}}^a_{\mathrm{GNSS}} &= \hat{\boldsymbol{e}} = \frac{\hat{z} \times \hat{\boldsymbol{u}}}{\|\hat{z} \times \hat{\boldsymbol{u}}\|} \\ \hat{\boldsymbol{r}}^t_{\mathrm{GNSS}} &= \hat{\boldsymbol{n}} = \frac{\hat{\boldsymbol{u}} \times (\hat{z} \times \hat{\boldsymbol{u}})}{\|\hat{\boldsymbol{u}} \times (\hat{z} \times \hat{\boldsymbol{u}})\|} \end{aligned}. \tag{28}$$

The receiver effective dipole vector can then be calculated as,

$$\boldsymbol{r}_{\mathrm{GNSS}} = \boldsymbol{P}\hat{\boldsymbol{e}} - \boldsymbol{S}_{\times}\hat{\boldsymbol{n}}. \tag{29}$$

The phase wind-up including both receiver feed rotation and satellite feed rotation for an observing RHCP GNSS antenna in the full crossed dipole formulation of Beyerle (2009) is given by,

$$\tan \Phi_{\mathrm{GNSS}} = \frac{\hat{\boldsymbol{e}}^T \boldsymbol{P}\hat{\boldsymbol{t}}^t + \hat{\boldsymbol{n}}^T \boldsymbol{P}\hat{\boldsymbol{t}}^a}{\hat{\boldsymbol{e}}^T \boldsymbol{P}\hat{\boldsymbol{t}}^a - \hat{\boldsymbol{n}}^T \boldsymbol{P}\hat{\boldsymbol{t}}^t}. \tag{30}$$

When computing all phase wind-up corrections detailed here, the boresight vectors for the receiver and the transmitter as well as the source unit vector $\hat{s}$ must be expressed in the same reference frame, either Earth-fixed or inertial. Assuming the transmitter boresight is antiparallel to the source unit vector, we can alternatively use the Wu et al. (1993) form:

$$\tan \Phi_{\mathrm{GNSS}} = \frac{[\boldsymbol{P}\hat{\boldsymbol{e}} - \boldsymbol{S}_{\times}\hat{\boldsymbol{n}}]^T \boldsymbol{S}_{\times}^T \boldsymbol{t}}{[\boldsymbol{P}\hat{\boldsymbol{e}} - \boldsymbol{S}_{\times}\hat{\boldsymbol{n}}]^T \boldsymbol{t}} = \frac{\hat{\boldsymbol{e}}^T \boldsymbol{S}_{\times}^T \boldsymbol{t} + \hat{\boldsymbol{n}}^T \boldsymbol{t}}{\hat{\boldsymbol{e}}^T \boldsymbol{t} - \hat{\boldsymbol{n}}^T \boldsymbol{S}_{\times}^T \boldsymbol{t}}. \tag{31}$$

Using the Wu et al. (1993) formalism, the total differential feed rotation can be expressed as a sum of two phase rotations, the first a rotation of the receiving feed about the pointing vector $\hat{s}$ relative to the north celestial pole, and the second a rotation of the satellite transmitting antenna about the pointing vector with respect to that same vector:

$$\begin{aligned}\Phi_{\mathrm{GNSS}} = \Phi_{\mathrm{sat}} + \Phi_{\mathrm{GNSS,\,rx}} &= \operatorname{atan}\frac{\hat{\boldsymbol{z}}^T \boldsymbol{S}_{\times}^T \boldsymbol{t}}{\hat{\boldsymbol{z}}^T \boldsymbol{t}} \\ &+ \operatorname{atan}\frac{\hat{\boldsymbol{e}}^T \boldsymbol{S}_{\times}^T \hat{\boldsymbol{z}} + \hat{\boldsymbol{n}}^T \boldsymbol{P}\hat{\boldsymbol{z}}}{\hat{\boldsymbol{e}}^T \boldsymbol{P}\hat{\boldsymbol{z}} - \hat{\boldsymbol{n}}^T \boldsymbol{S}_{\times}^T \hat{\boldsymbol{z}}}.\end{aligned} \tag{32}$$

Note that Equations 31 and 32 may differ by $\pm 2\pi$ radians, so cycle number adjustment may be necessary for consistency between the two forms. As the GNSS antenna does not slew to observe a source, the receiver feed rotation is caused only by the rotation of Earth in an inertial frame. Interestingly, this split expression also allows us to find the correction needed for GNSS antenna observations of distant natural radio sources such as AGNs. Neglecting any change in the intrinsic polarization of the observed natural radio source, the term $\Phi_{\mathrm{sat}} = 0$; thus, the feed rotation is simply,

$$\Phi_{\mathrm{GNSS,\,nat}} = \operatorname{atan}\frac{\hat{\boldsymbol{e}}^T \boldsymbol{S}_{\times}^T \hat{\boldsymbol{z}} + \hat{\boldsymbol{n}}^T \boldsymbol{P}\hat{\boldsymbol{z}}}{\hat{\boldsymbol{e}}^T \boldsymbol{P}\hat{\boldsymbol{z}} - \hat{\boldsymbol{n}}^T \boldsymbol{S}_{\times}^T \hat{\boldsymbol{z}}}. \tag{33}$$

This correction is needed for observations of natural radio sources with GNSS antennas such as those demonstrated with a GNSS antenna to radio telescope interferometer in Skeens et al. (2023). It also may be relevant for the stationary dipole antennas that comprise the low-frequency array (LOFAR) (van Haarlem et al. 2013).

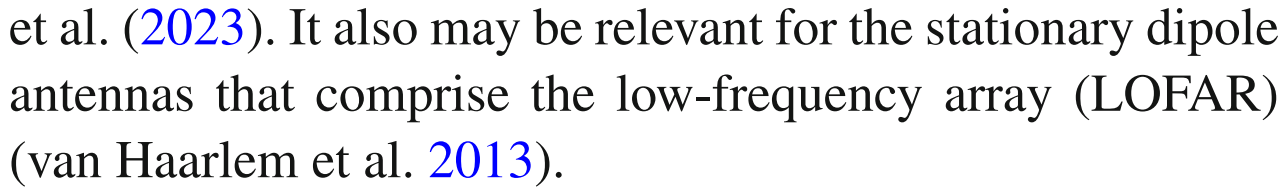

To track the feed rotation of the radio telescope, we must define a vector primitive that follows the orientation of the feed horn of the telescope as it slews to observe a source in the direction $\hat{s}$. For a radio telescope observing in the RHCP basis, the boresight, aligned, and transverse dipole vectors can be defined as,

$$\begin{aligned}\hat{\boldsymbol{r}}_{\mathrm{tel}}^b &= \hat{\boldsymbol{s}} \\ \hat{\boldsymbol{r}}_{\mathrm{tel}}^a &= \frac{(\hat{\boldsymbol{s}} \times \hat{\boldsymbol{a}}) \times \hat{\boldsymbol{s}}}{\|(\hat{\boldsymbol{s}} \times \hat{\boldsymbol{a}}) \times \hat{\boldsymbol{s}}\|} = \frac{\boldsymbol{P}\hat{\boldsymbol{a}}}{\|\boldsymbol{P}\hat{\boldsymbol{a}}\|}. \\ \hat{\boldsymbol{r}}_{\mathrm{tel}}^t &= \frac{\hat{\boldsymbol{s}} \times \hat{\boldsymbol{a}}}{\|\hat{\boldsymbol{s}} \times \hat{\boldsymbol{a}}\|} = \frac{\boldsymbol{S}_{\times}\hat{\boldsymbol{a}}}{\|\boldsymbol{S}_{\times}\hat{\boldsymbol{a}}\|}\end{aligned} \tag{34}$$

As in Subsection 2.2, we can simply use the projection of the fixed-axis vector as the receiver effective dipole, $\boldsymbol{r} = \boldsymbol{P}\hat{\boldsymbol{a}}$. The full phase wind-up for an observing RHCP radio telescope is therefore given by,

$$\tan \Phi_{\mathrm{tel}} = \frac{\hat{\boldsymbol{a}}^T \boldsymbol{S}_{\times}^T \boldsymbol{t}}{\hat{\boldsymbol{a}}^T \boldsymbol{t}}. \tag{35}$$

Splitting the differential feed rotation into receiver and satellite terms:

$$\Phi_{\mathrm{tel}} = \operatorname{atan}\frac{\hat{\boldsymbol{z}}^T \boldsymbol{S}_{\times}^T \boldsymbol{t}}{\hat{\boldsymbol{z}}^T \boldsymbol{t}} + \operatorname{atan}\frac{\hat{\boldsymbol{a}}^T \boldsymbol{S}_{\times}^T \hat{\boldsymbol{z}}}{\hat{\boldsymbol{a}}^T \boldsymbol{P}\hat{\boldsymbol{z}}}. \tag{36}$$

Note that the receiver feed rotation angle is equivalent to the receiver feed rotation given in the context of natural radio source observations in Equation 13. For an equatorial or polar mount, this feed rotation correction is $\Phi_{\mathrm{rx}} = 0$. Thus, the full differential feed rotation correction for this mount type is simply given by,

$$\Phi_{\mathrm{pol}} = \Phi_{\mathrm{sat}} = \operatorname{atan}\frac{\hat{\boldsymbol{z}}^T \boldsymbol{S}_{\times}^T \boldsymbol{t}}{\hat{\boldsymbol{z}}^T \boldsymbol{t}}. \tag{37}$$

For an FN focus, Dodson (2009) give the feed rotation correction for natural radio source observations in terms of the parallactic angle $\Psi$ as,

$$\begin{aligned}\Phi_{\mathrm{FN,\,rx}} &= \Psi \pm E \\ &= \operatorname{atan}\frac{\hat{\boldsymbol{u}}^T \boldsymbol{S}_{\times}^T \hat{\boldsymbol{z}}}{\hat{\boldsymbol{u}}^T \boldsymbol{P}\hat{\boldsymbol{z}}} \pm \operatorname{asin}(\hat{\boldsymbol{u}}^T \hat{\boldsymbol{s}}),\end{aligned} \tag{38}$$

where the sign of the elevation angle, $E$, added to the original correction is determined by whether the reflection from the third mirror is right-handed (+) or left-handed (−). The BWG model (Dodson and Rioja 2022) has an additional

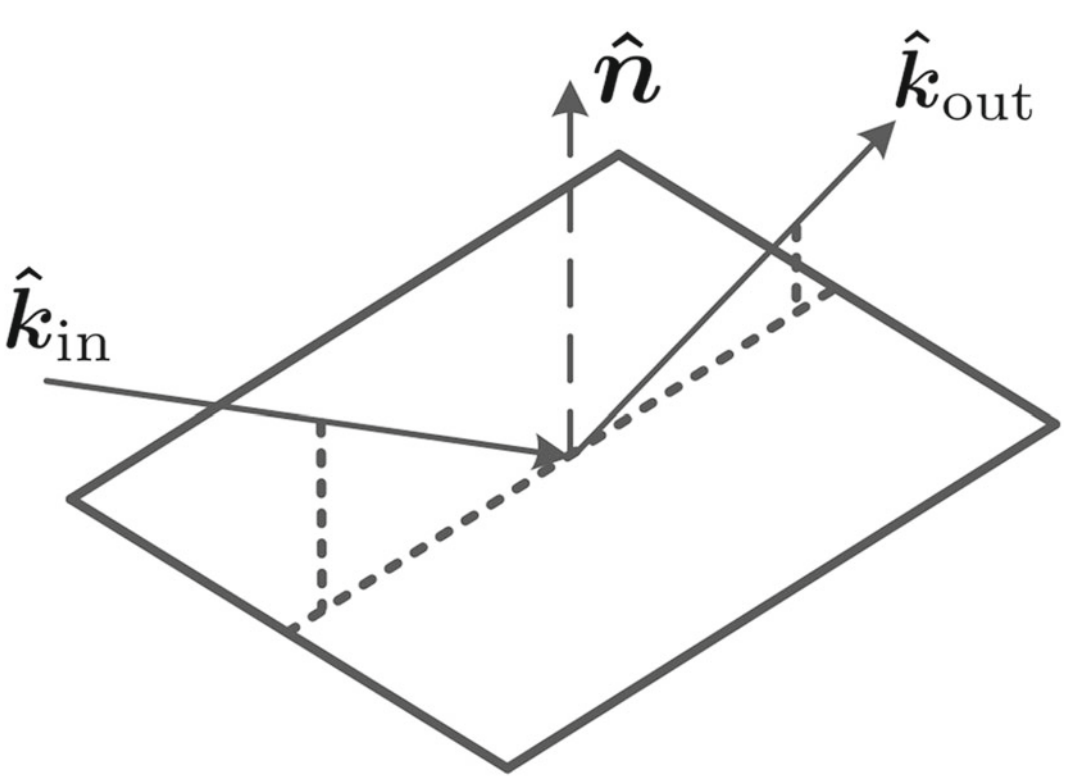

**Fig. 6** A reflection of a photon with incoming wave vector $\hat{\boldsymbol{k}}_{\mathrm{in}}$ and outgoing wave vector $\hat{\boldsymbol{k}}_{\mathrm{out}}$ from a mirror with normal vector, $\hat{\boldsymbol{n}}$

azimuth dependence because the focus is fixed to the stationary antenna support structure:

$$\begin{aligned}\Phi_{\mathrm{BWG,\,rx}} &= \Psi \pm E \mp A \\ &= \operatorname{atan}\frac{\hat{\boldsymbol{u}}^T \boldsymbol{S}_{\times}^T \hat{\boldsymbol{z}}}{\hat{\boldsymbol{u}}^T \boldsymbol{P} \hat{\boldsymbol{z}}} \pm \operatorname{asin}(\hat{\boldsymbol{u}}^T \hat{\boldsymbol{s}}) \mp \operatorname{atan}\frac{\hat{\boldsymbol{e}}^T \hat{\boldsymbol{s}}}{\hat{\boldsymbol{n}}^T \hat{\boldsymbol{s}}}.\end{aligned} \tag{39}$$

The sign of the azimuth dependence, $A$, is again determined by the handedness of a mirror reflection. For satellite observations, we need to add the satellite rotations to these expressions:

$$\Phi_{\mathrm{FN}} = \Phi_{\mathrm{FN,\,rx}} + \Phi_{\mathrm{sat}}, \tag{40}$$

$$\Phi_{\mathrm{BWG}} = \Phi_{\mathrm{BWG,\,rx}} + \Phi_{\mathrm{sat}}. \tag{41}$$

To rigorously demonstrate the accuracy of these expressions, we explicitly consider the effect of reflections through the waveguide of a radio telescope. This method of tracking reflections can also be used to find the differential feed rotation for an arbitrary future waveguide configuration.

### 3.2 Tracking the polarization orientation through reflections

For an incoming wave vector representing the propagation direction of a photon, $\hat{\boldsymbol{k}}_{\mathrm{in}}$, and an outgoing wave vector, $\hat{\boldsymbol{k}}_{\mathrm{out}}$, we can define a normal to the reflecting surface as,

$$\hat{\boldsymbol{n}} = \frac{\hat{\boldsymbol{k}}_{\mathrm{out}} - \hat{\boldsymbol{k}}_{\mathrm{in}}}{\|\hat{\boldsymbol{k}}_{\mathrm{out}} - \hat{\boldsymbol{k}}_{\mathrm{in}}\|}. \tag{42}$$

Figure 6 shows the incoming and outgoing wave vectors for a reflecting surface with normal vector $\hat{\boldsymbol{n}}$.

In Beyerle (2009), the effect of reflection on the aligned and transverse components of the incoming electric field from the transmitter is considered explicitly. We will adapt this model to use the simpler Wu et al. (1993) formalism, as the two approaches are equivalent for all radio telescopes that track observed sources. In this case, we need only reflect the effective transmitter dipole across the surface normal. When reflected, the phase angle flips; thus, the direction of the reflected vector is negated:

$$\boldsymbol{q} = -(\boldsymbol{t} - 2[\boldsymbol{t}^T \hat{\boldsymbol{n}}]\hat{\boldsymbol{n}}). \tag{43}$$

For the FN and BWG focus types, we must consider multiple reflections from mirrors with different orientations. Equation 43 can be used to find the aligned and transverse components of the electric field after reflection $j + 1$ by replacing the initial transmitter effective dipole vector, $\boldsymbol{t}$, with the effective dipole vector after reflection $j$:

$$\boldsymbol{q}_{j+1} = -(\boldsymbol{q}_j - 2[\boldsymbol{q}_j^T \hat{\boldsymbol{n}}]\hat{\boldsymbol{n}}). \tag{44}$$

By applying Equations 43 and 44 for all mirrors in the waveguide, we can rigorously find the transmitted electric field observed at the feed horn within the receiving radio telescope. The differential feed rotation correction can then be computed using the Wu et al. (1993) formalism as,

$$\Phi = \operatorname{atan}\frac{\hat{\boldsymbol{r}}^a \cdot (\boldsymbol{q} \times \hat{\boldsymbol{k}}_{\mathrm{out}}) + \hat{\boldsymbol{r}}^t \cdot \boldsymbol{q}}{\hat{\boldsymbol{r}}^a \cdot \boldsymbol{q} - \hat{\boldsymbol{r}}^t \cdot (\boldsymbol{q} \times \hat{\boldsymbol{k}}_{\mathrm{out}})}. \tag{45}$$

In Appendix C, we show the reflections as they occur in two real radio telescopes—the FN focus YEBES40M (Appendix C.1) and the BWG focus WARK30M (Appendix C.2). As noted in Dodson and Rioja (2022), at the correlation step during VLBI processing, the detected LHCP or RHCP polarizations are relabeled to represent the true 'on-sky' value. This means that if an odd number of reflections occurs in the waveguide, the phase is flipped an additional 180 degrees as if it had been reflected from one more perfect mirror. To compensate for this, simply take the negative of the correction in Equation 45.

### 3.3 Verifying the full Nasmyth and beam waveguide feed rotation models with dual polarization observations

For telescopes with dual circularly polarized receivers observing a weakly polarized source, the feed rotation in the right-handed (RR) and left-handed (LL) visibilities is equivalent but of opposite sign. Thus, if the RHCP and LHCP visibilities are fringe fit separately, the feed rotation effect can be directly measured by differencing the RHCP and LHCP fringe phases. Upon differencing the fringe phases, the phase rotation due to geometric delay, the delay caused by the neutral atmosphere, and the dispersive delay from the ionosphere cancel. This leaves the differential feed rotation, a small delay due to the birefringent property of the ionosphere, the delay caused by instrumental effects at antennas 1 and 2, and the additive noise expected in fringe fitting. The delay from birefringence is assumed to be small, and the instrumental delay is generally slowly varying, meaning that the phase difference is to first order characterized by twice the differential feed rotation with an additive offset from instrumental effects:

$$\begin{aligned}\phi^R - \phi^L &= 2\pi f\left(\tau_{\text{iono, bi}} + \tau_{\text{instr}}\right) + (\Phi_2 - \Phi_1) \\ &\quad - (-\Phi_2 + \Phi_1) \approx 2\left(\Phi_2 - \Phi_1\right) + \Phi_{\text{instr}}\end{aligned}. \tag{46}$$

We can use Equation 46 to examine the VLBI feed rotation model for the FN and BWG mount types by observing natural radio sources in dual polarization. Geodetic VLBI experiments rarely record both the right-handed and left-handed data, so astronomical experiments are the best source of dual polarization measurements to do this analysis. Most astronomical experiments include many observations of bright and weakly polarized AGNs located near the source of scientific interest that serve as phase and amplitude calibrators. These calibrator scans have high signal-to-noise ratio and are good targets to examine the phase difference between RHCP and LHCP.

First, we process data from the C band experiment EB050 (Bondi et al. 2016), available from the archive of the Joint Institute for VLBI in Europe (JIVE) and recorded in 2011. These data include visibilities from baselines with the European VLBI Network (EVN) telescope YEBES40M in central Spain, which has the full Nasmyth focus type. We examine the baseline YEBES40M–EFLSBERG, where EFLSBEG is a 100-m-diameter azimuth–elevation radio telescope with a Gregorian focus, meaning its feed rotation is simply given by Equation 15. We fringe fitted these visibilities in the software PIMA (Petrov et al. 2011), and we used the full reflection-based model detailed in Appendix C.1 to model the feed rotation correction. Comparing the reflection-based feed rotation correction to Equation 38, the models are exactly identical in this case.

Figure 7 shows on the left the difference of the right-handed and left-handed fringe phase for the baseline EFLSBERG–YEBES40M. The phases have been unwrapped to the same cycle as the corresponding point in the differenced feed rotation model, and the differenced feed rotation model on the right side of the figure has been shifted by a best-fit constant offset to account for the additive offset shown in Equation 46. Different sources are plotted in the figure with different colors. The modeled and observed feed rotation agree quite well with some additional scatter in the observed fringe phases.

Figure 8 shows the same differenced fringe phase plotted against azimuth (left) and elevation (right) rather than time. The fringe phase is plotted in units of degrees rather than radians to give a 1:1 scale with azimuth and elevation expressed in their typical units. This allows us to see the strong dependence of the feed rotation correction on elevation. Here, the differenced phase and feed rotation model are plotted on the same axes, showing the correspondence directly.

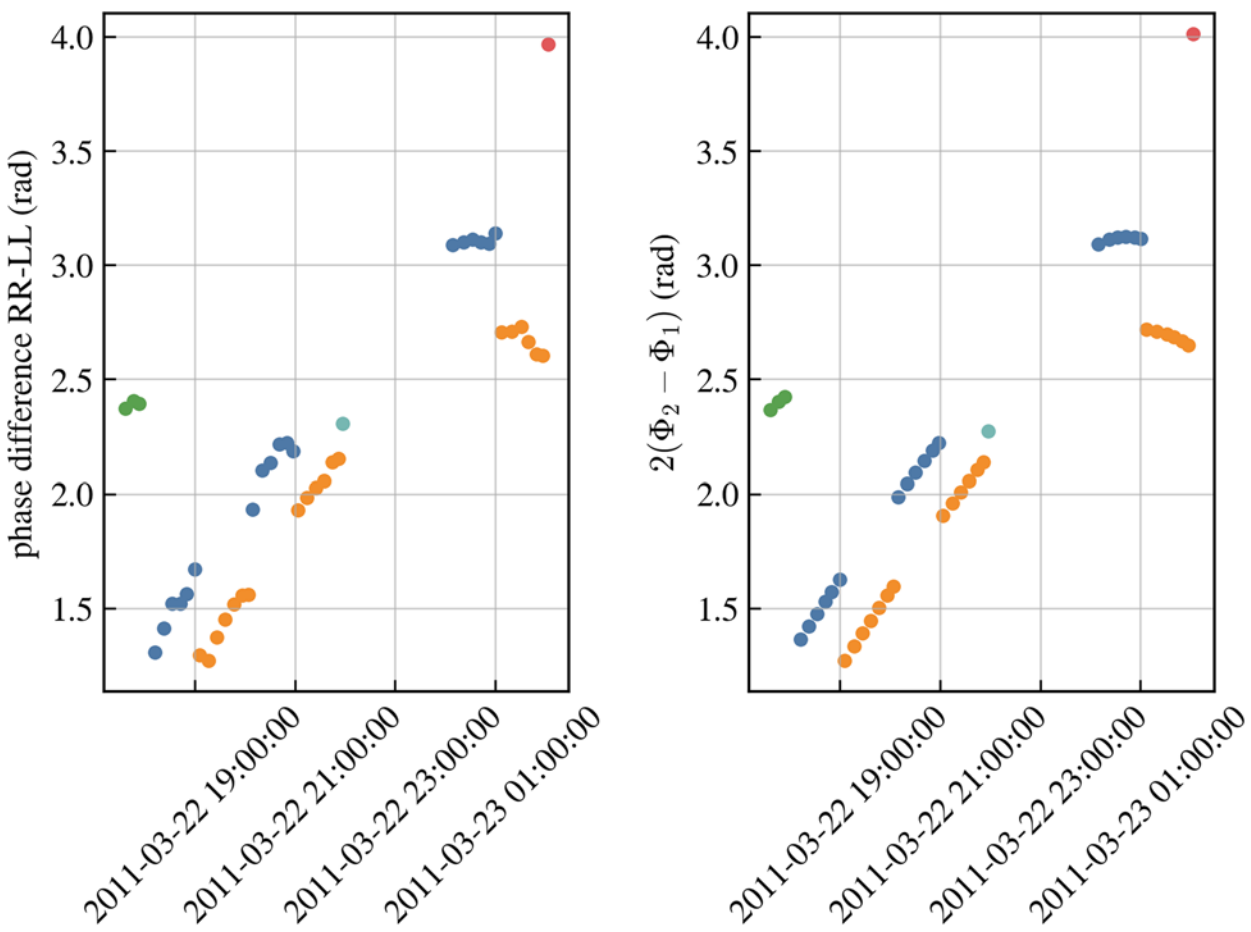

**Fig. 7** Difference of the RR and LL fringe phases on the baseline EFLSBEG–YEBES40M (left) and the predicted effect of feed rotation on the RR–LL phase difference (right). Different radio sources are plotted in different colors

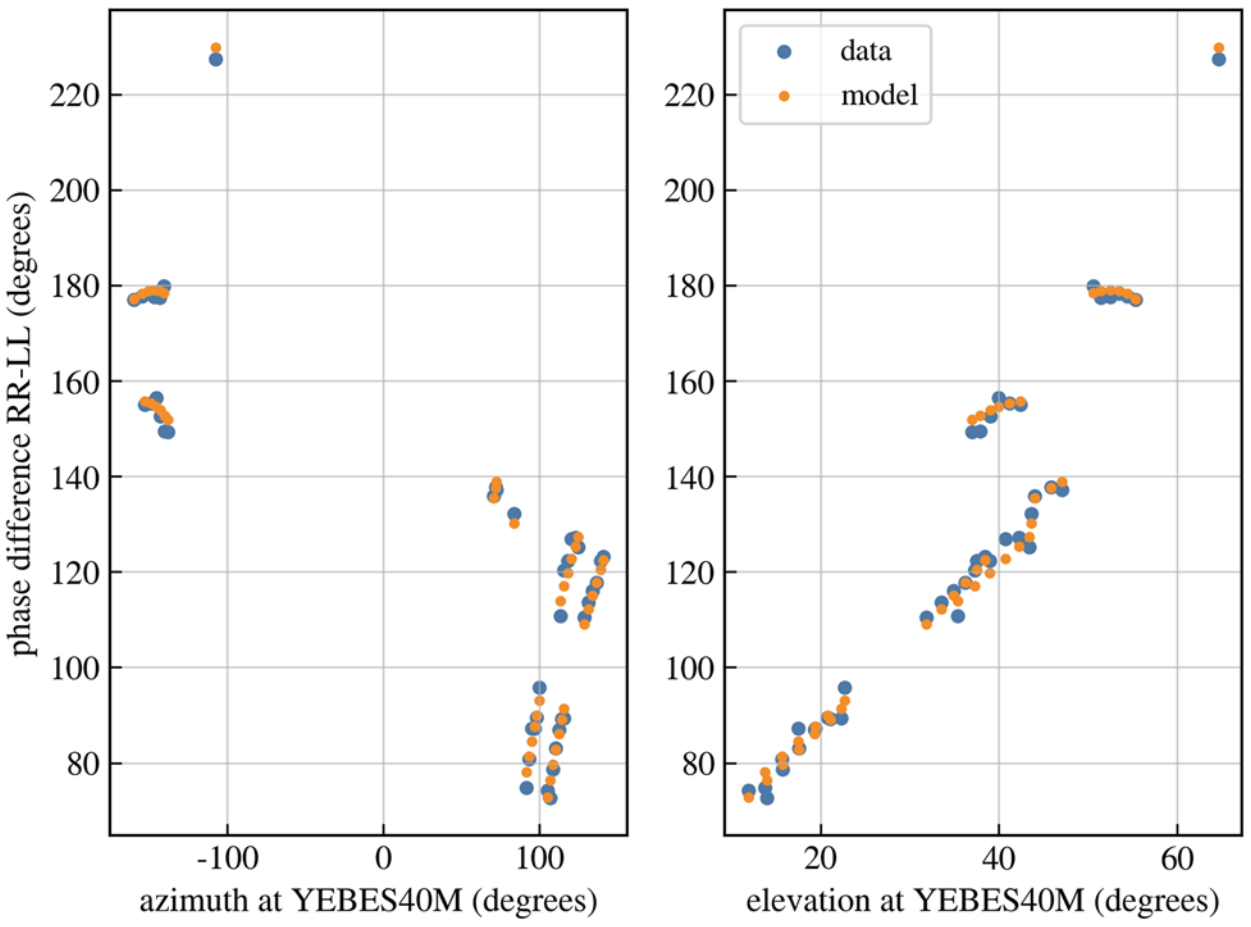

**Fig. 8** EFLSBERG–YEBES40M empirical phase difference between the right- and left-handed fringe phases as a function of azimuth (left) and elevation (right) and the same phase differences as predicted by the feed rotation model

To test the accuracy of the BWG model, we have processed the X band JIVE experiment GM074Z, which contains visibilities from baselines including WARK30M, an azimuth–elevation mounted, BWG focus radio telescope in Warkworth, New Zealand. We examine the baseline WARK30M–PARKES, where PARKES is a 64-m telescope with a typical azimuth–elevation mount and a prime focus.

The data are plotted against azimuth (left) and elevation (right) in Fig. 9. The model given in Dodson and Rioja (2022) and shown in Equation 39 describes the feed rotation correc-tion as the parallactic angle correction with an additional linear dependence on the elevation angle and azimuth angle. This agrees exactly with the reflection-based model given in Appendix C.2.

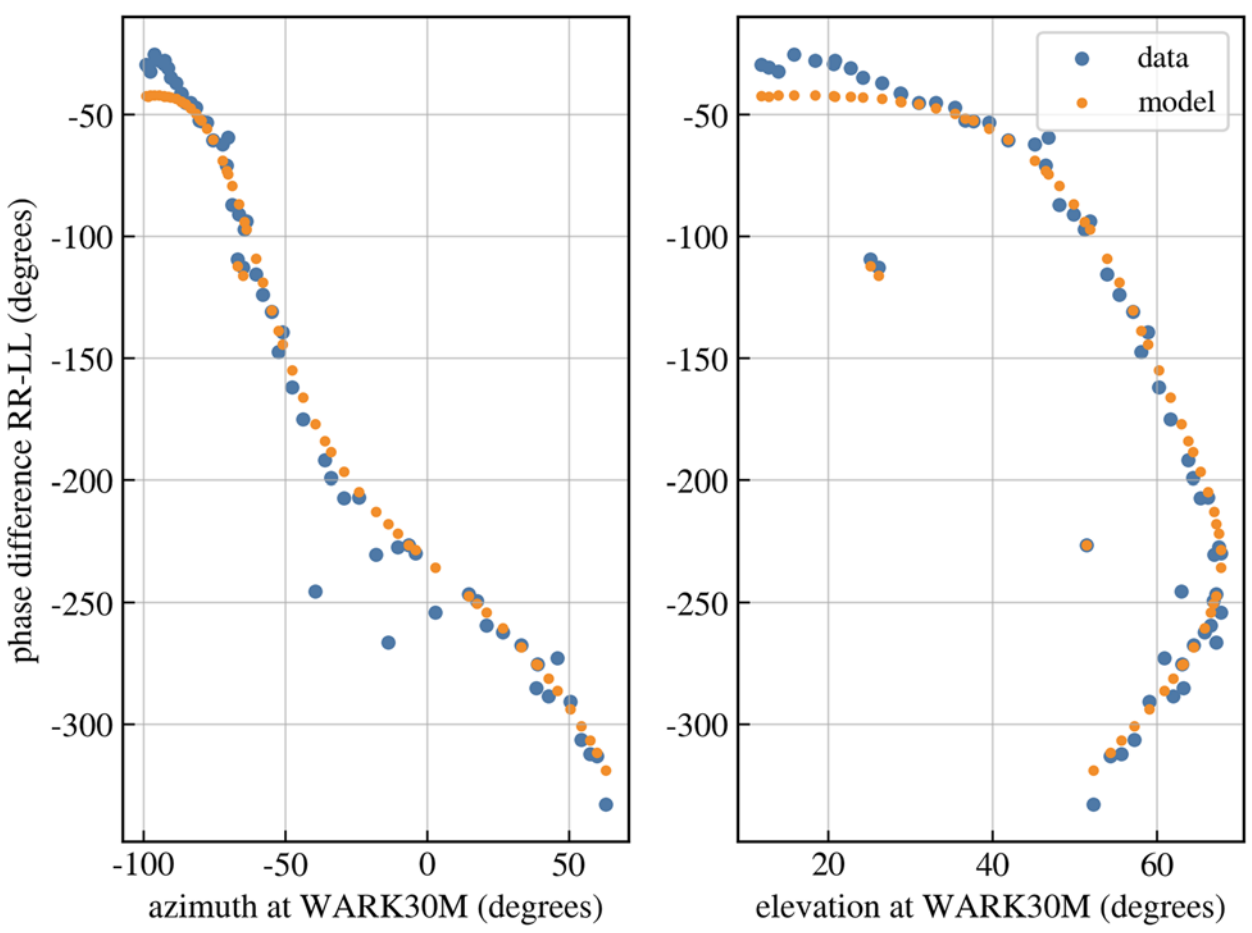

**Fig. 9** PARKES–WARK30M empirical phase difference between the right- and left-handed fringe phases as a function of azimuth (left) and elevation (right) and the same phase differences as predicted by the feed rotation model

## 4 Simulating the differential feed rotation effect in satellite observations

### 4.1 Differential feed rotation in a short baseline satellite observation

To show the magnitude of the differential feed rotation effect and the breakdown of the phase rotation due to the satellite and receiver feed rotation, we have conducted a simple two-line element (TLE)-based simulation of a satellite pass over a mixed GNSS antenna–azimuth–elevation radio telescope baseline. The positions we used are the locations of a deployed choke ring GNSS antenna (DBR205) and the Fort Davis Very Long Baseline Array (FD-VLBA) radio telescope during an experiment conducted on January 25, 2023. We simulated Global Positioning System (GPS) Satellite Vehicle Number (SVN) 76, which was observed in this experiment. The positions of the two antennas are given in Table 1, and the TLE orbit for the simulated satellite is shown in Table 2 and was obtained from CelesTrak. Further details about this experiment and its goals for generating local tie vectors are available in Skeens et al. (2024). Here, we will focus on the differential feed rotation effect on the 73.5 m baseline between DBR205 and FD-VLBA.

For each of the two antennas, Fig. 10 shows the feed rotation of the satellite along the line-of-sight direction and the feed rotation of the receiver as described in Equations 21 and 22. While the magnitude of the phase rotation caused by the changing attitude of the satellite is large, it is common to both antennas, as the short baseline means that the line-of-sight vector from the receiver to the satellite, $\hat{\boldsymbol{s}}$, is nearly identical. Because this satellite feed rotation is common, it will be removed in differential measurements such as phase delays. However, for longer baselines the line-of-sight vector can change significantly. This is particularly true for observations of satellites with lower orbital altitude.

The feed rotation of the receiver is quite different between the two antennas, leading to a large discrepancy in the total observed differential feed rotation. The GNSS antenna does not move in the Earth-fixed frame, and as such, its apparent feed rotation is caused only by the movement of the satellite and the rotation of Earth. In contrast, the radio telescope tracks the satellite actively and for that reason has a much larger feed rotation effect due to a combination of Earth rotation and slewing.

In VLBI processing, what will be observed is the difference of the two differential feed rotations, or the differential phase wind-up between the two antennas. Figure 11 shows at the top the total differential feed rotation for each antenna and on the bottom the differential feed rotation delay, labeled rotation delay, that would be observed due to these differential feed rotations. The delay is presented in time units rather than cycles assuming the observation is taken at the L1 frequency, 1575.42 MHz: $\tau_{\mathrm{dfr}} = \frac{\Phi_2 - \Phi_1}{2\pi f}$. The phase ambiguity interval corresponding to this frequency is $\Delta\tau = 634.8$ ps. For the two antennas, the resultant delay is at the level of hundreds of picoseconds—a significant fraction of this ambiguity interval—and therefore must be corrected to obtain an accurate differential positioning solution with phase delays.

### 4.2 Differential feed rotation in a long baseline satellite observation

In a long baseline satellite observation, the difference in the line of sight from each antenna to the satellite and therefore the direction of the source unit vector means that the changing orientation of the satellite will cause an observable change in differential phase measurements. A significant upcoming application for long baseline VLBI observations of satellites is the GENESIS mission, whose purpose is to colocate all four space geodetic techniques onboard a satellite. The precise details of the orbit this satellite will be placed in are not publicly available, but Delva et al. (2023) describe the planned orbit type as a 6000 km altitude, quasi-polar circular orbit. To show the likely magnitude of the feed rotation correction for this satellite, we have simulated a simple possible orbit with a right ascension of the ascending node of 30 deg, a quasi-polar inclination of 97 deg, and no drag or other non-gravitational effects. We again propagate this orbit using SGP4 as a two-line element, arbitrarily selecting the same date as the first simulation. No pointing model is available for this satellite, so we use the simple model defined in Equation 26 that points the transmitting antenna at the Earth center while keeping solar panels oriented toward the Sun. We set the observing telescopes as two VGOS stations, GGAO12M at the Goddard Geophysical and Astronomical Observatory (GGAO) and WETTZ13S at Geodetic Observatory Wettzell. The positions and mount types of the telescopes are given in Table 3.

**Table 1** ITRF positions and mount types for the simulated radio antennas

| Antenna | Type | X (m) | Y (m) | Z (m) |
|---|---|---|---|---|
| DBR205 | GNSS | −1324070.478 | −5332176.001 | 3231921.799 |
| FD-VLBA | Az.-El | −1324009.454 | −5332181.955 | 3231962.369 |

**Table 2** Two-line element for GPS Satellite Vehicle Number 76 obtained from CelesTrak for the date January 25, 2023

| | | | | | | | | |
|---|---|---|---|---|---|---|---|---|
| 1 | 41328U | 16007A | 23024.39332179 | .00000015 | 00000-0 | 00000+0 | 0 | 9995 |
| 2 | 41328 | 54.9577 | 138.7050 | 0066950 | 231.2234 | 128.1638 | 2.00560135 | 51012 |

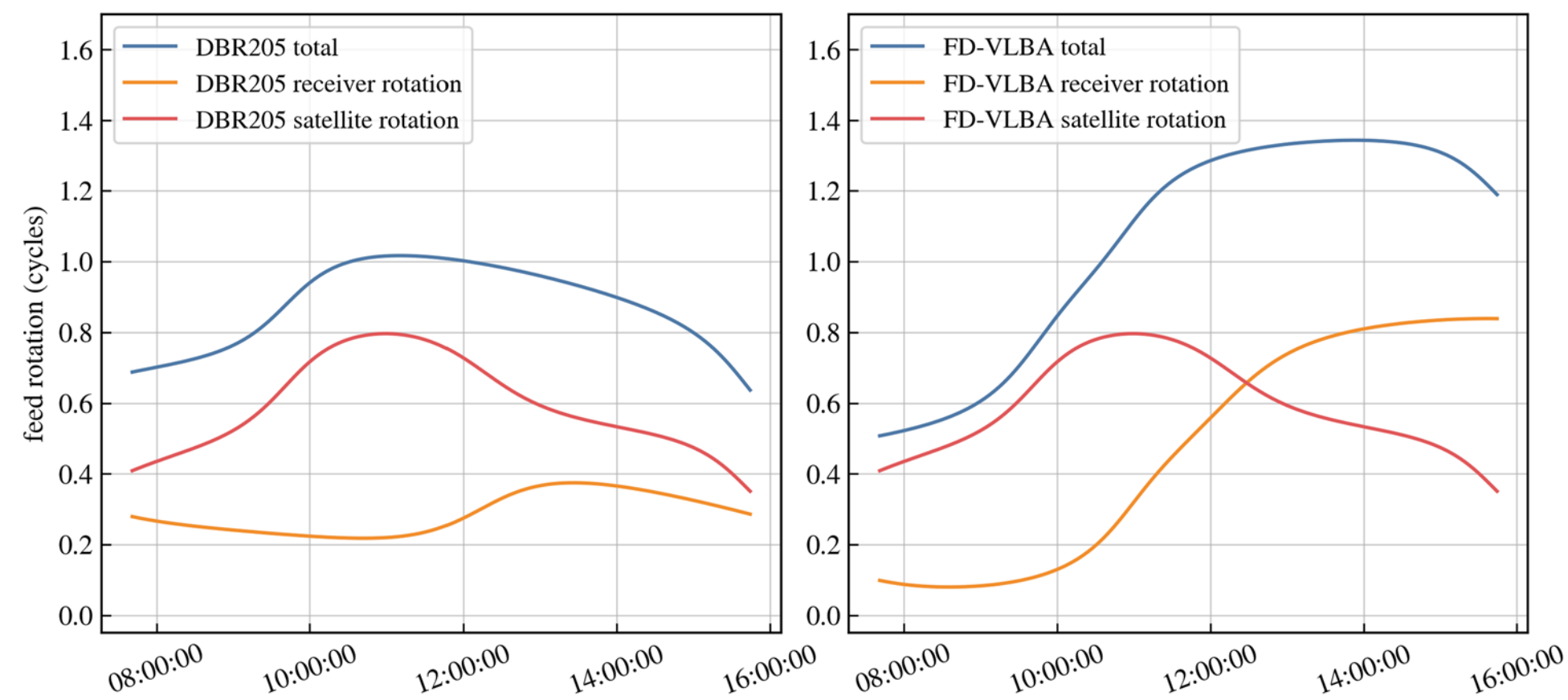

**Fig. 10** Feed rotation or phase wind-up due to receiver and satellite attitude changes for the GNSS antenna DBR205 and the azimuth–elevation telescope FD-VLBA

Many of the next generation radio telescopes, such as those designed to meet VGOS standards including GGAO12M and WETTZ13S, use dual linearly polarized receivers because the systems are designed to observe across wide bandwidths (2-14 GHz), and circular polarizers have diminished performance far from their design frequencies (Marti-Vidal et al. 2016). The most common method of processing dual linearly polarized observations for geodetic VLBI is combining the orthogonal linear polarizations in the so-called pseudo-Stokes I combination as described in Cappallo (2014). This methodology assumes that the typically processed natural radio sources are weakly polarized. However, the GENESIS satellite may have an LHCP VLBI transmitter (Karatekin et al. 2023). Thus, the phase corrections needed to correctly combine visibilities in the horizontal, vertical, and cross-hand polarizations after correlation while ensuring phase and amplitude coherence to form Stokes I will be very complicated. As the vast majority of the received power will be LHCP, it may be possible to combine the baseband streams from the horizontal and vertical channels for each antenna into a single left-handed stream, thereby vastly reducing the correlation time, minimizing the amount of complex visibility data, and simplifying the analysis.

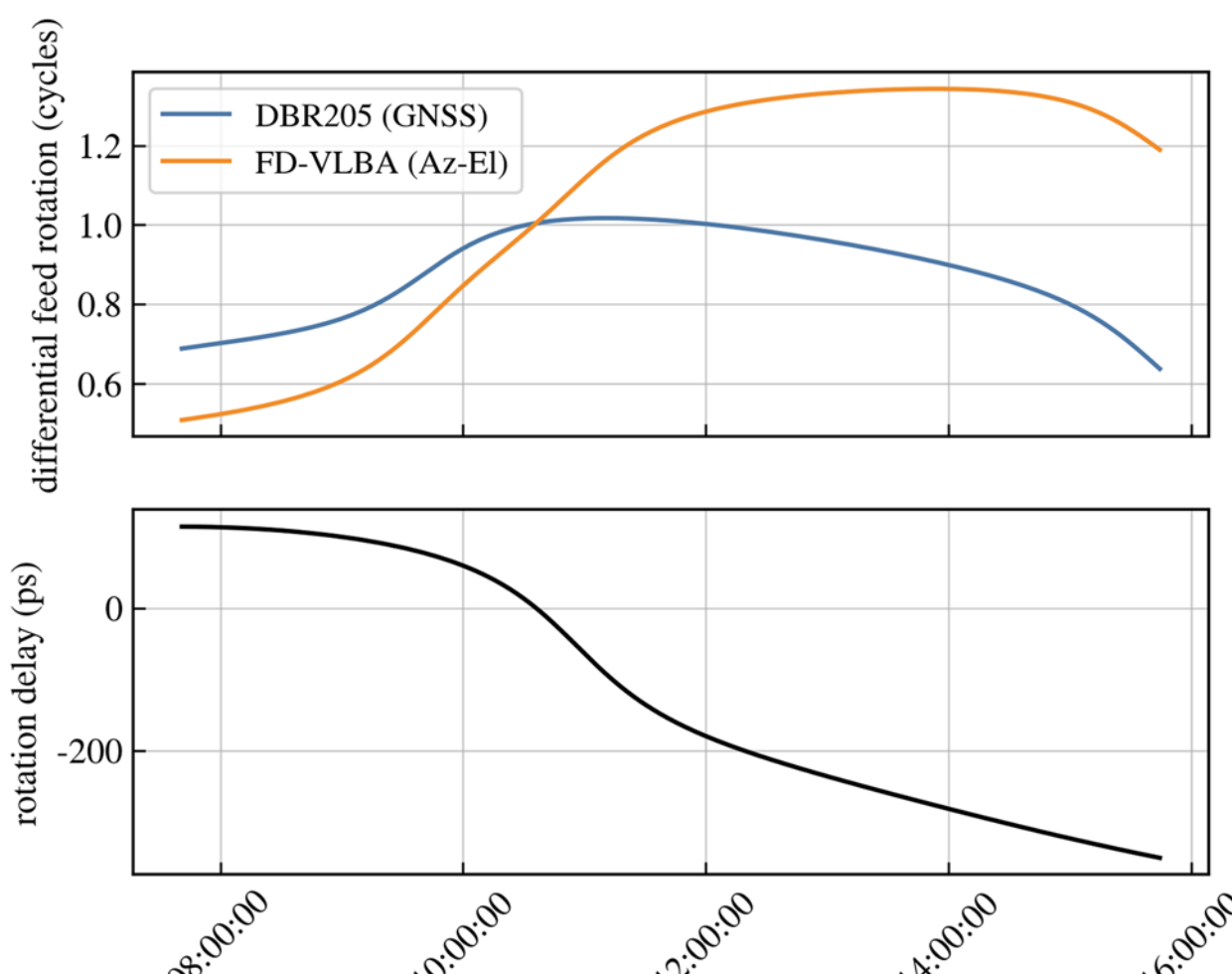

**Fig. 11** Total differential feed rotation for GNSS antenna DBR205 and Az-El radio telescope FD-VLBA and the feed rotation or wind-up delay between them

For an LHCP transmitter and receiver, the differential feed rotation correction is given by the negative of the RHCP–RHCP correction. This LHCP correction is plotted in Fig.

**Table 3** ITRF positions and mount types for two VLBI Global Observing System (VGOS) telescopes

| Antenna | Type | X (m) | Y (m) | Z (m) |
|---|---|---|---|---|
| GGAO12M | Az.-El | 1130730.179 | −4831245.954 | 3994228.241 |
| WETTZ13S | Az.-El | 4075659.180 | 931824.551 | 4801516.102 |

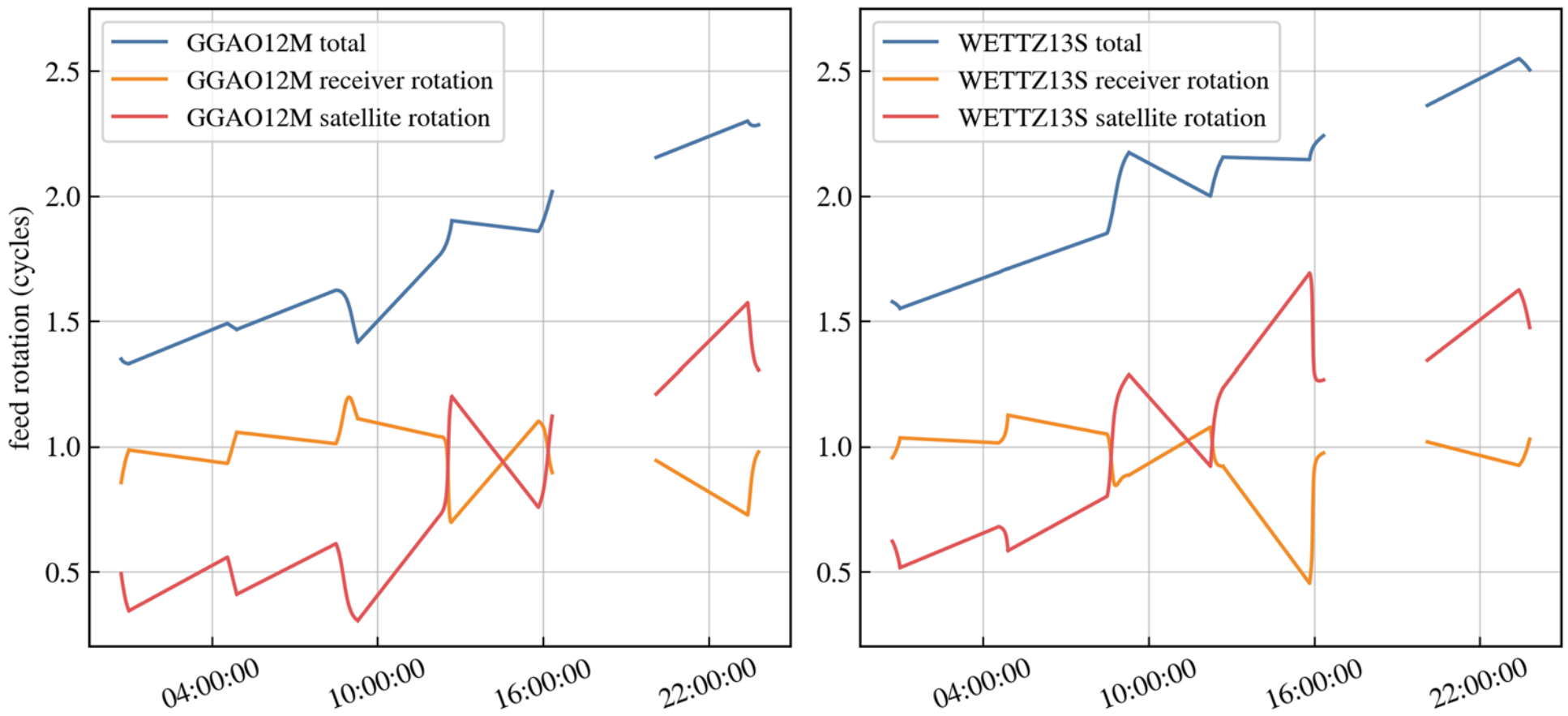

**Fig. 12** Feed rotation due to receiver and satellite attitude changes for the Az-El radio telescopes GGAO12M and WETTZ13S observing the GENESIS satellite

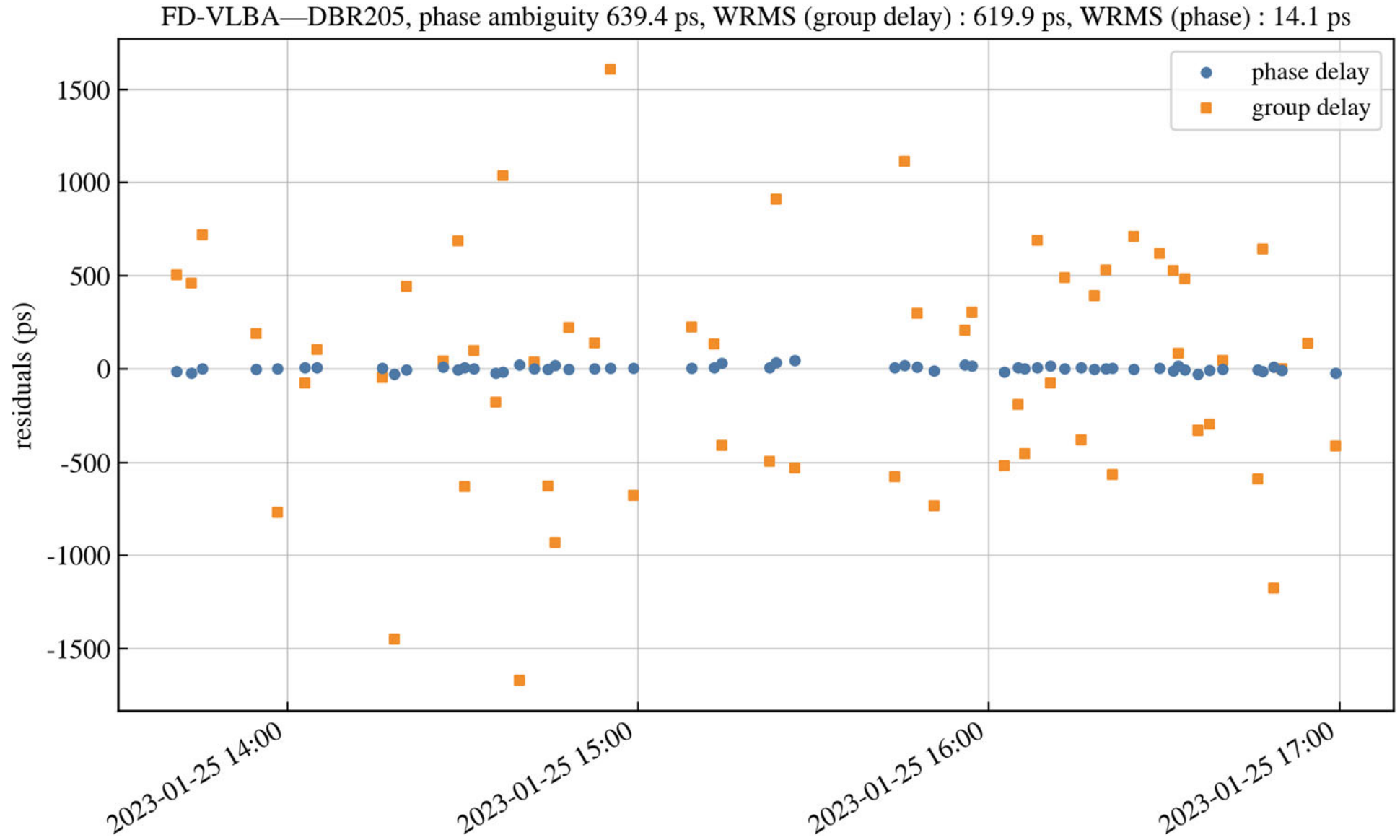

**Fig. 13** Postfit residuals for the group delay and phase delay solutions with GNSS satellite sources

12, which shows the receiver feed rotation, satellite feed rotation, and total feed rotation observed by each telescope. In this simulation, unlike in the short baseline simulation, the observed phase shift due to changing satellite attitude is noticeably different between the two observing telescopes. This discrepancy in the observed satellite feed rotation is caused by the difference in the look direction from the receiving antenna to the transmitting antenna, $\hat{s}$. The difference of the satellite feed rotations leads to a large and observable feed rotation delay that would not be compensated using a model that does not consider the orientation of the transmitting antenna.

We should note that when processing VLBI data from linearly polarized receivers or from mixed linearly and circularly polarized receivers, one must first compute an instrumental complex bandpass model using data from calibrator sources prior to processing data from target sources. When processing calibrator sources, the complex visibilities must be corrected for the feed rotation $\Phi$ of receivers, i.e., multiplying them by the complex phasor $(\cos \Delta\Phi), i\,(\sin \Delta\Phi)$ with an appropriate sign. After the complex bandpass model is computed and applied to the complex visibilities of the target sources, the visibility data can be transformed to the Stokes I, Q, U, V bases, to the XX, XY, YX, YY bases, or to the RR, RL, LR, LL bases for further analysis in a preferred set of bases. Transformation from the original bases to the preferred bases will again require adjusting the visibility data by the feed rotation of the receivers. Parallactic angle and feed rotation are othen used interchangeably in the literature or user manuals (for instance, Thompson et al. (2017); Greisen (2003); Marti-Vidal et al. (2016)), but we should stress that the feed rotation is only equivalent to the parallactic angle for telescopes with an azimuth–elevation mount and a feed horn located in the primary, Gregorian, or Cassegrain focus.

**Table 4** Baseline vector components and associated uncertainties in millimeters with the additive uncertainty in picoseconds need to achieve chi2pdof of unity for the baseline DBR205–FD-VLBA

| Observable | $E$ | $\sigma_E$ | $N$ | $\sigma_N$ | $U$ | $\sigma_U$ | $L$ | $\sigma_L$ | $\sigma_{\text{add}}(ps)$ |
|---|---|---|---|---|---|---|---|---|---|
| Group delay | 39435.5 | 59.1 | 60639.8 | 43.4 | 13040.0 | 116.0 | 73501.0 | 48.9 | 647.1 |
| Phase delay | 39447.4 | 1.3 | 60630.2 | 1.0 | 12996.4 | 2.7 | 73491.7 | 1.1 | 14.7 |

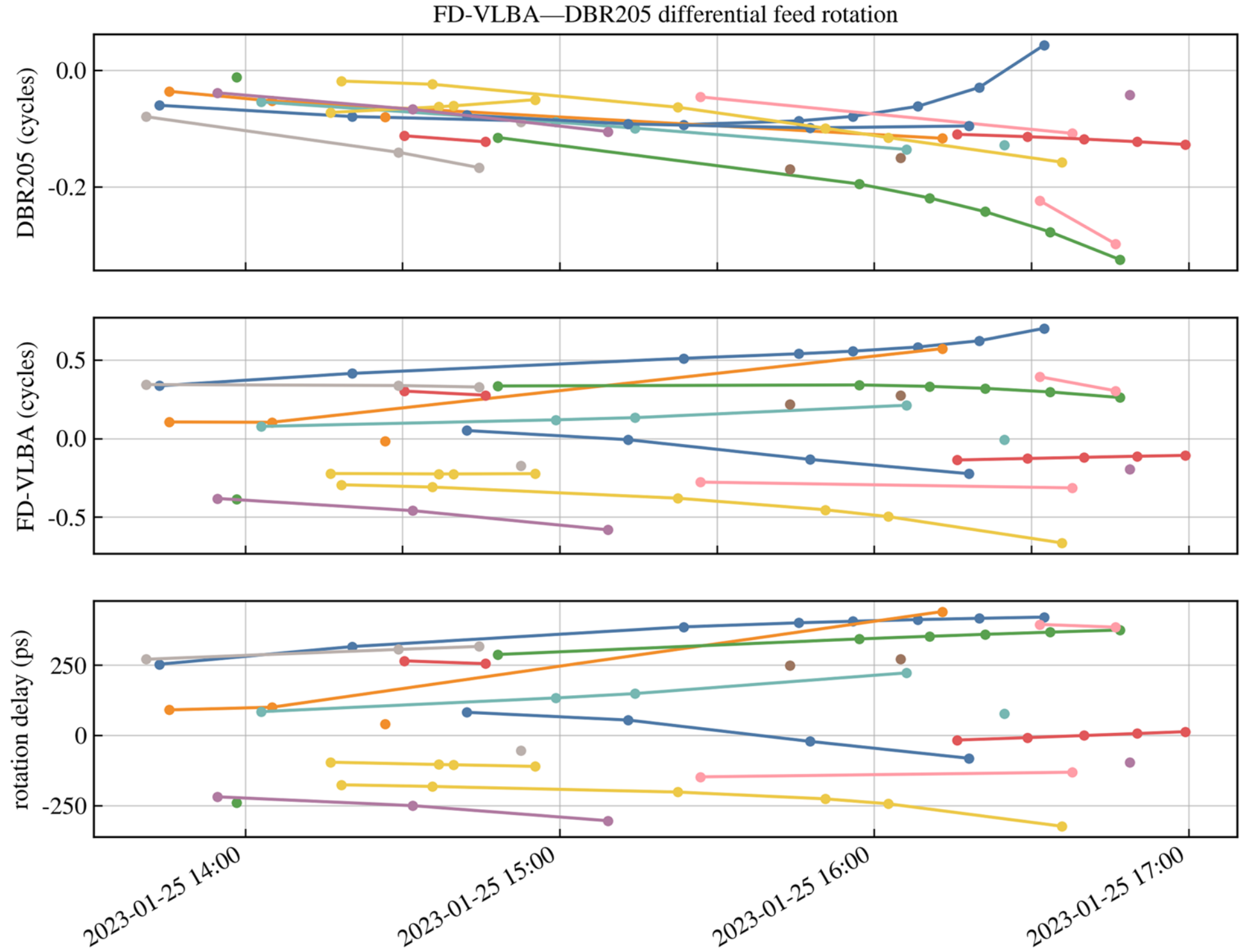

**Fig. 14** Differential feed rotation between satellite and receiver in cycles at the GNSS antenna DBR205 and the radio telescope FD-VLBA (top, middle) and the corresponding delay signal caused by the difference in feed rotation (bottom). Points connected by a line indicate observations of the same satellite

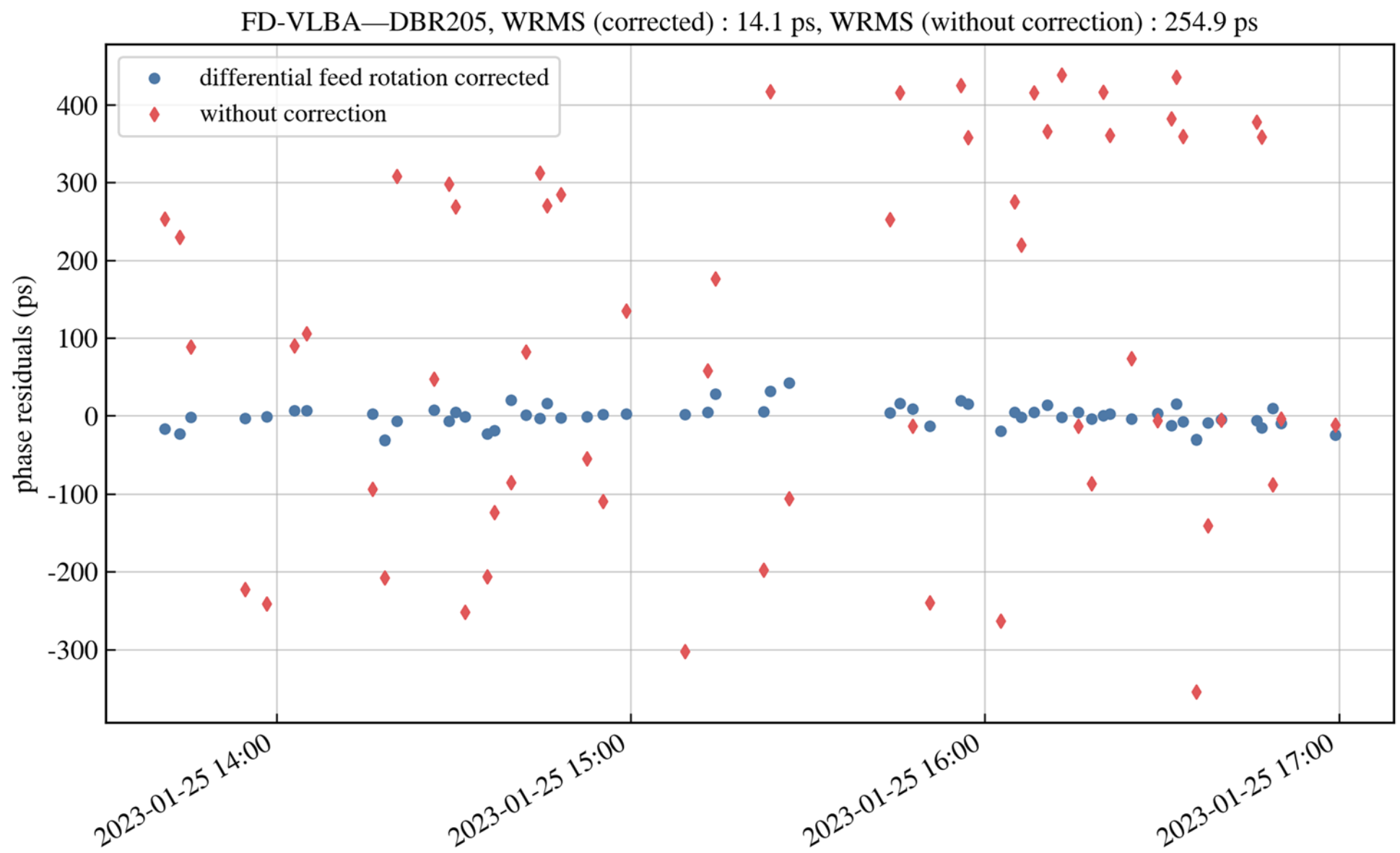

**Fig. 15** Phase delay postfit residuals with and without the differential feed rotation correction on baseline DBR205–FD-VLBA

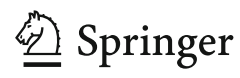

## 5 Analysis of short baseline geodetic VLBI experiments using phase delays

Because the differential feed rotation effect induces a phase change that is independent of frequency, it does not appear in group delay measurements of a single polarization. However, it is critical to account for the differential feed rotation to obtain accurate phase measurements. This is particularly important when the feed rotations of the observing telescopes are very different, meaning that the feed rotation signal does not cancel. To demonstrate the validity of the described models, we will show the results of two VLBI experiments that depend on accurate feed rotation modeling to produce high-quality phase delay solutions. Both of the analyzed experiments recorded data only in RHCP.

### 5.1 GNSS satellite observations with a GNSS antenna–radio telescope interferometer

First, we analyze an experiment conducted on January 25, 2023, between the GNSS antenna DBR205 and the radio telescope FD-VLBA. The data shown here comprise 60 observations of 20 satellites from the GPS, Galileo, and BeiDou systems collected over about 3.5 h. As described in Skeens et al. (2024), a third GNSS antenna placed about 9 km from the two analyzed here participated in this experiment. Here we are interested in displaying only the differential feed rotation effect rather than linking the GNSS and VLBI reference frames, so we focus on the short baseline at which the highest quality data were collected. For each of the 60 observations or scans of about 30 s length, we obtain one group delay and one phase delay measurement after correlation in DiFX (Deller et al. 2007, 2011) and fringe fitting in PIMA. The near-field delay model used to correlate the baseband samples from the GNSS receiver and radio telescope is adapted from Jaron and Nothnagel (2019) and is described fully along with partial derivatives for station positions in Skeens and Petrov (2024). DBR205 was connected to the same hydrogen maser frequency standard used by FD-VLBA to minimize error caused by clock drift. We use a least squares estimator in this analysis with a single linear clock function covering the full experiment. DBR205 is set as the reference station for both position and clock parameters, and we do not estimate any atmospheric parameters. The satellite orbits are those estimated by the Center for Orbit Determination in Europe (CODE) in their final multi-GNSS precise ephemerides, and the satellite attitudes are modeled using the COD ORBEX files. Phase Center Offset (PCO) and Phase Center Variation (PCV) are accounted for in the GNSS satellites and the Topcon CR-G5 GNSS antenna using the IGS ANTEX file.

As is the case with all phase delay VLBI analysis, we first compute a group delay-only solution. After the group delay solution, we resolve the phase delay integer cycle ambiguity by correcting the phase delay cycle number such that the measurements form a single continuous time series with a root-mean-square difference much smaller than the ambiguity interval. After resolving the phase delay ambiguity, we compute a second phase delay-only positioning solution using the much more precise phase delay measurements. In the phase delay solution, correctly accounting for the differential feed rotation delay is critical. Figure 13 shows the postfit residuals for the group delay-only and phase delay-only positioning solutions. Because the GNSS bands are only 10-20 MHz wide, the precision of group delay measurements collected from their emission is relatively low. The phase delay measurement precision is inversely proportional to the sky frequency regardless of bandwidth, so once phase delay ambiguities are resolved, these measurements can be used to obtain a much more precise position.

Table 4 shows the components of the baseline vector in millimeters with associated formal uncertainties in east, north, up coordinates. As is standard in VLBI processing, we use a diagonal weight matrix of the correlation uncertainties for the group and phase delays and iteratively fit a constant variance term, $\sigma^2_{\mathrm{add}}$, such that the Chi-squared per degree of freedom ($\frac{\chi^2}{\nu}$) is unity. The additive uncertainty fit to the measurements in each solution is shown in units of time in the table. As expected, these additive uncertainties are of the same order of magnitude of the weighted root-mean-square residuals in Fig. 13. The correlation and fringe fitting errors modeled by the fringe fitting software are orders of magnitude smaller and are thus insignificant compared to the true uncertainties of the measurements.

Each observed satellite has a unique time series of differential feed rotation for each antenna. Figure 14 shows the differential feed rotation in cycles for each antenna and satellite, where the modeled feed rotations corresponding to observations of the same satellite are connected by a line and different satellites are plotted with different colors. Colors repeat when the number of satellites plotted exceeds the length of the color cycle, but new satellites with the same color as those previously plotted are not connected by a line. At the bottom of the figure, a third plot shows the rotation/wind-up delay caused by the difference of the differential feed rotations at each antenna. As discussed pre-viously, this rotation delay for each satellite is large due to the differences in antenna orientation. To produce a multi-satellite positioning solution using phase delays, it is critical to account for these rotation delays.

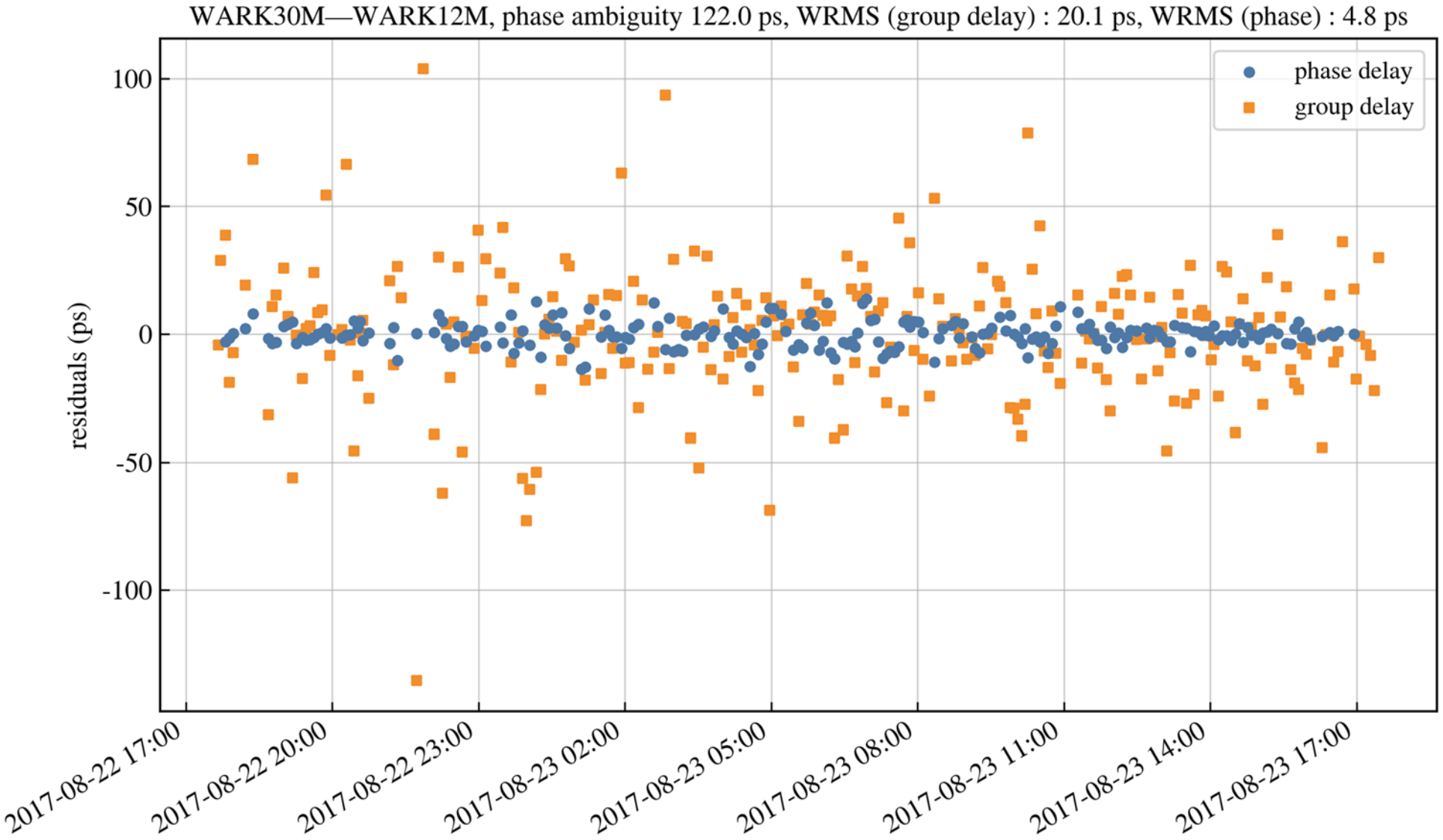

**Fig. 16** Postfit residuals for the group delay and phase delay solutions on the baseline WARK30M–WARK12M

**Table 5** Baseline vector components and associated uncertainties in millimeters with the additive uncertainty in picoseconds need to achieve a reduced Chi-squared of unity for the baseline WARK30M–WARK12M

| Observable | $E$ | $\sigma_E$ | $N$ | $\sigma_N$ | $U$ | $\sigma_U$ | $L$ | $\sigma_L$ | $\sigma_{\text{add}}(ps)$ |
|---|---|---|---|---|---|---|---|---|---|
| Group delay | 183108.0 | 5.6 | −27382.1 | 4.8 | −5546.1 | 15.9 | 185227.1 | 4.4 | 16.8 |
| Phase delay | 183104.8 | 1.4 | −27382.2 | 1.2 | −5532.8 | 4.5 | 185223.5 | 1.1 | 5.1 |

Finally, Fig. 15 shows the phase delay times series with and without the differential feed rotation delay correction. It is important to note that the uncorrected residuals shown here are not computed after a secondary least squares adjustment, they are simply found by removing the differential feed rotation from the postfit phase delay residuals directly. Were a new positioning solution computed without the correction, much of the error seen in these uncorrected residuals would instead be projected into the state space, biasing the estimated position and thereby lowering the observed WRMS. This is precisely why applying the correction is pivotally important. We show the phase delay postfit residuals and tie vector estimated while omitting the differential feed rotation correction in Appendix E.

## 5.2 A first local tie vector for baseline WARK30M–WARK12M

Using the feed rotation model for radio telescopes with BWG focuses, we have processed a short baseline experiment in X band (about 8.33 GHz) between the BWG telescope WARK30M and the Cassegrain azimuth–elevation telescope WARK12M located about 183 m away at the same observatory in Warkworth, New Zealand. These data are from the Southern Astronomy Project geodetic VLBI campaign, and we analyze in particular the experiment AUA026 conducted on August 23, 2017. A group delay tie vector has previously been shown for this baseline in Petrov et al. (2015), which describes the details of the WARK30M telescope design in detail. This is again a 24-hour experiment, and we used a first degree B-spline (piecewise linear) clock function with 60-minute node spacing to model the differential clock variation. We also estimated a B-spline of first degree for the differential Zenith Wet Delay (dZWD) with 240-minute node spacing. As with the satellite observation experiment, we first compute a group delay-only solution and resolve the integer ambiguities of the phase delays by comparison to the group delay model. After fixing the phase delays to the correct ambiguities, we compute a least squares solution for the phase delay measurements. The ambiguity interval of these X band observations is much smaller at about 122 ps. WARK12M was set as the reference for position, clock, and dZWD estimation.

Figure 16 shows the postfit residuals for group delays and phase delays. The effective RMS bandwidth of this experiment is much larger and the frequency higher than that of the GNSS satellite observations, so the WRMS of the group delay residuals is much lower––21.0 ps here as opposed to 620 ps for the satellite observations. The postfit residuals

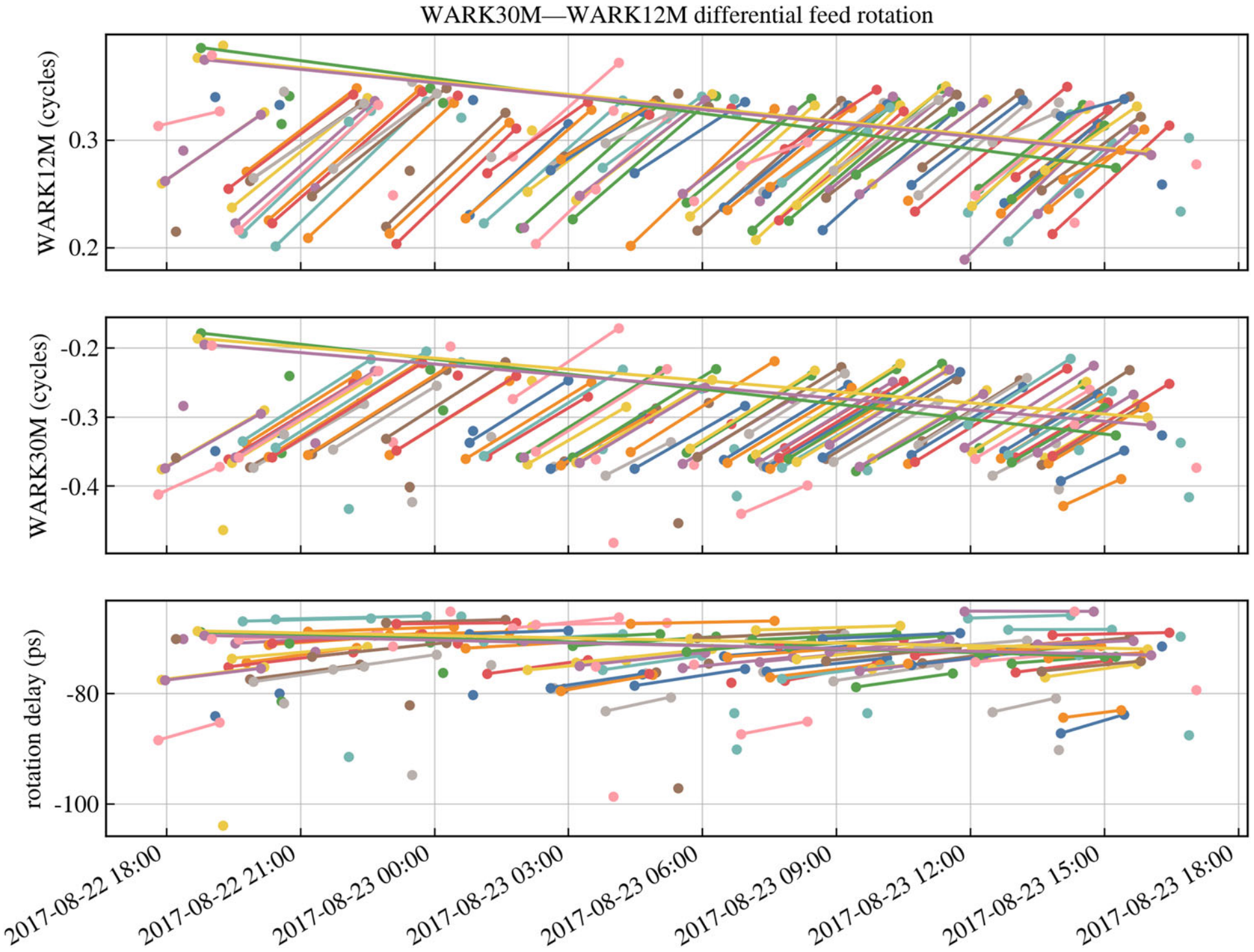

**Fig. 17** Feed rotation in cycles at the Cassegrain focus WARK12M and the beam waveguide (BWG) telescope WARK30M (top, middle) and the corresponding delay signal caused by the difference in feed rotation (bottom)

show no noticeable structure, and the phase delay WRMS is about 4.8 ps.

Table 5 shows the components of the WARK30M–WARK12M baseline vector in millimeters with associated formal uncertainties in east, north, up coordinates. Because the group delays are much more precise, the group delay solution is of much higher quality than the corresponding GNSS group delay solution, at a similar level of precision to the phase delay solution. The group and phase delay baseline vectors agree to within about 1 standard deviation, and the uncertainties of the phase delay solution are significantly reduced. This indicates that the feed rotation model is performing well, and the estimated tie vector is accurate.

Finally, Fig. 17 demonstrates the magnitude of the modeled feed rotation in both radio telescopes. Because both radio telescopes have azimuth–elevation mounts and are close together, the large majority of the feed rotation difference in the rotation delay is caused by the additional elevation and azimuth dependence in WARK30M due to its BWG focus.

## 6 Conclusions

We have presented a unified model of differential feed rotation that works for both GNSS antennas and radio telescopes observing either satellites or natural radio sources. The model has been explicitly laid out for each of the extant radio telescope mount types and focus types. This model can be decomposed into two phase rotations—a rotation of the transmitter about the north celestial pole projected along the line-of-sight vector and a rotation of the receiver feed tracked against the projection of the north celestial pole vector. The receiver feed rotation term in this decomposition is equivalent to the existing correction for natural radio sources such as AGNs. Using the two rotation decomposition, we identified the feed rotation correction for a stationary GNSS antenna observing natural radio sources. We have also examined the feed rotation effect for radio telescopes with the full Nasmyth and beam waveguide focus types. These focus types have additional reflections in the radio telescope structure, which cause complicated phase rotations. By explicitly considering each of the reflections, we have developed a rigorous model for the feed rotation of the FN focus telescope YEBES40M and the BWG focus telescope WARK30M. We use the differing sign of the feed rotation effect in dual circular polarization observations to measure the magnitude of the feed rotation effect in two astronomical experiments, EB050 and GM074Z, in which YEBES40M and WARK30M participated, respectively. We find that the reflection-based models match those described in Dodson (2009) and Dodson and Rioja (2022) exactly for the case of RHCP observations of natural radio sources.

To illustrate the importance of modeling the differential feed rotation correction, we performed simple simulations of satellite observations, propagating the TLE orbits with a Python implementation (Rhodes 2023) of SGP4 as described in Vallado and Crawford (2008). On a short GNSS antenna–azimuth–elevation radio telescope baseline observing a GPS satellite, the contribution of the receiver feed rotation is a significant fraction of the phase ambiguity interval. While the satellite feed rotation contribution to the phase measured at each antenna is large, the common line of sight of the satellite on a short baseline means that the effect mostly cancels in differential measurements. This is not the case for the second simulation of radio telescope observations (GGAO12M–WETTZ13S) of the upcoming GENESIS satellite using orbital parameters described in Delva et al. (2023). The difference in the line-of-sight vectors due to the large distance between the observing telescopes and the lower orbital altitude leads to a non-negligible contribution to the visibility phase from the changing attitude of the observed satellite.

We have analyzed two geodetic VLBI experiments to demonstrate the feed rotation model and its effectiveness in correcting phase measurements. The first experiment is a VLBI co-observation of GNSS satellites with a choke ring GNSS antenna and azimuth–elevation radio telescope (FD-VLBA). Using phase delays to obtain an accurate positioning solution is critical in observations of GNSS satellites because the relatively small bandwidth of the signals emitted by the satellites means that group delay measurements have much lower precision. The phase delay WRMS error is a factor of 44 times lower than the group delay WRMS in this experiment. After resolving the integer cycle ambiguity and correcting the differential feed rotation effect, we are able to estimate a baseline vector between the two antennas with formal errors of 1-3 mm in east, north, up coordinates with only 3.5 h of data.

The second analyzed experiment is the baseline WARK30M–WARK12M in the Southern Astronomy Project experiment AUA026. WARK30M and WARK12M both use an azimuth–elevation mount, but WARK30M has a BWG focus. We use the reflection-based BWG feed rotation model detailed in this work to correct the phase delay measurements. With this correction, we were able to resolve the phase delay ambiguities. Using the fixed-ambiguity phase measurements, we have produced the first published phase delay tie vector for this baseline.

It is not uncommon in the literature, including textbooks and user manuals, to use the term parallactic angle in place of the more general term feed rotation when covering the most common case of radio telescopes with azimuth–elevation mounts. However, the parallactic angle is equivalent to the feed rotation angle only for a limited number of cases. Some antennas, for instance YEBES40M or WARK30M, can be viewed as azimuth–elevation when considering them as a mechanical structure, but as their feed horn is not located in the primary, Gregorian, or Cassegrain focus but rather in a cabin that does not fully corotate, their feed rotation is different than other azimuth–elevation telescopes like EFFLSBRG. The use of parallactic angle as a proxy for feed rotation angle is also not applicable for observing Earth-orbiting satellites. This is especially important when data from linearly polarized feeds are processed, because an error in the computation of the feed rotation may prevent correct complex band calibration. We hope our article will prompt re-examination of the existing code used for data analysis, and we hope that the use of the term parallactic angle in place of feed rotation will be avoided in the future.

## Appendix A Derivation of the geometrically idealized model from the crossed dipole model

Starting from Equation 5, we first assume that $\hat{\boldsymbol{r}}^b = \hat{\boldsymbol{s}}$; therefore, $\hat{\boldsymbol{r}}^t = \boldsymbol{S}_\times \hat{\boldsymbol{r}}^a$. The receiver effective dipole is then given by,

$$\begin{aligned} \boldsymbol{r} &= \boldsymbol{P}\hat{\boldsymbol{r}}^a - \boldsymbol{S}_\times(\boldsymbol{S}_\times \hat{\boldsymbol{r}}^a) = \boldsymbol{P}\hat{\boldsymbol{r}}^a + \boldsymbol{P}\hat{\boldsymbol{r}}^a \\ &= 2\boldsymbol{P}\hat{\boldsymbol{r}}^a. \end{aligned} \tag{A1}$$

Because the magnitude of $\boldsymbol{r}$ will cancel in the wind-up expressions, we can simplify this to $\boldsymbol{r} = \boldsymbol{P}\hat{\boldsymbol{r}}^a$. The model is then given by,

$$\begin{aligned} \tan\Phi &= \frac{[\hat{\boldsymbol{r}}^a]^T \boldsymbol{P}\hat{\boldsymbol{t}}^t + [\boldsymbol{S}_\times \hat{\boldsymbol{r}}^a]^T \boldsymbol{P}\hat{\boldsymbol{t}}^a}{[\hat{\boldsymbol{r}}^a]^T \boldsymbol{P}\hat{\boldsymbol{t}}^a - [\boldsymbol{S}_\times \hat{\boldsymbol{r}}^a]^T \boldsymbol{P}\hat{\boldsymbol{t}}^t} \\ &= \frac{[\hat{\boldsymbol{r}}^a]^T \left(\boldsymbol{S}_\times^T \hat{\boldsymbol{t}}^a + \boldsymbol{P}\hat{\boldsymbol{t}}^t\right)}{[\hat{\boldsymbol{r}}^a]^T \left(\boldsymbol{P}\hat{\boldsymbol{t}}^a - \boldsymbol{S}_\times^T \hat{\boldsymbol{t}}^t\right)}. \end{aligned} \tag{A2}$$

Because $\boldsymbol{P}$ is idempotent and symmetric, and $\boldsymbol{S}_\times^T \boldsymbol{P} = \boldsymbol{P}\boldsymbol{S}_\times^T = \boldsymbol{S}_\times^T$, we can show the equivalence to Equation 7 as,

$$\begin{aligned} \tan\Phi &= \frac{[\hat{\boldsymbol{r}}^a \boldsymbol{P}]^T \boldsymbol{S}_\times^T \left(\boldsymbol{P}\hat{\boldsymbol{t}}^a + \boldsymbol{S}_\times \hat{\boldsymbol{t}}^t\right)}{[\hat{\boldsymbol{r}}^a \boldsymbol{P}]^T \left(\boldsymbol{P}\hat{\boldsymbol{t}}^a + \boldsymbol{S}_\times \hat{\boldsymbol{t}}^t\right)} \\ &= \frac{\boldsymbol{r}^T \boldsymbol{S}_\times^T \boldsymbol{t}}{\boldsymbol{r}^T \boldsymbol{t}}. \end{aligned} \tag{A3}$$

Starting again from Equation 5, we relax the receiver boresight assumption and instead assume $\hat{\boldsymbol{t}}^b = -\hat{\boldsymbol{s}}$; therefore, $\hat{\boldsymbol{t}}^t = \boldsymbol{S}_\times^T \hat{\boldsymbol{t}}^a$. The transmitter effective dipole is then,

$$\begin{aligned} \boldsymbol{t} &= \boldsymbol{P}\hat{\boldsymbol{t}}^a + \boldsymbol{S}_\times(\boldsymbol{S}_\times^T \hat{\boldsymbol{t}}^a) = \boldsymbol{P}\hat{\boldsymbol{t}}^a + \boldsymbol{P}\hat{\boldsymbol{t}}^a \\ &= 2\boldsymbol{P}\hat{\boldsymbol{t}}^a. \end{aligned} \tag{A4}$$

Again, we take $\boldsymbol{t} = \boldsymbol{P}\hat{\boldsymbol{t}}^a$ for simplicity because the magnitude of the vector will cancel. Simplifying the model once more, we have,

$$
\begin{aligned}
\tan\Phi &= \frac{[\hat{\boldsymbol{r}}^a]^T \boldsymbol{P}[\boldsymbol{S}_\times^T \hat{\boldsymbol{t}}^a] + [\hat{\boldsymbol{r}}^a]^T \boldsymbol{P}\hat{\boldsymbol{t}}^a}{[\hat{\boldsymbol{r}}^a]^T \boldsymbol{P}\hat{\boldsymbol{t}}^a - [\hat{\boldsymbol{r}}^t]^T \boldsymbol{P}[\boldsymbol{S}_\times^T \hat{\boldsymbol{t}}^a]} \\
&= \frac{\left([\hat{\boldsymbol{r}}^a]^T \boldsymbol{S}_\times^T + [\hat{\boldsymbol{r}}^t]^T \boldsymbol{P}\right)\hat{\boldsymbol{t}}^a}{\left([\hat{\boldsymbol{r}}^a]^T \boldsymbol{P} - [\hat{\boldsymbol{r}}^t]^T \boldsymbol{S}_\times^T\right)\hat{\boldsymbol{t}}^a} \\
&= \frac{\left(\boldsymbol{P}\hat{\boldsymbol{r}}^a - \boldsymbol{S}_\times \hat{\boldsymbol{r}}^t\right)^T \boldsymbol{S}_\times^T (\boldsymbol{P}\hat{\boldsymbol{t}}^a)}{\left(\boldsymbol{P}\hat{\boldsymbol{r}}^a - \boldsymbol{S}_\times \hat{\boldsymbol{r}}^t\right)^T (\boldsymbol{P}\hat{\boldsymbol{t}}^a)} \\
&= \frac{\boldsymbol{r}^T \boldsymbol{S}_\times^T \boldsymbol{t}}{\boldsymbol{r}^T \boldsymbol{t}}.\square
\end{aligned}
\tag{A5}
$$

## Appendix B Derivation of the split phase wind-up expression

By inspection, we assume the transmitter rotation term by taking $\hat{\boldsymbol{r}} = \boldsymbol{P}\hat{z}$ for a fictitious RHCP receiver aligned with the celestial north pole. Simplifying Equation 7:

$$
\begin{aligned}
\tan\Phi_{\mathrm{tx}} &= \frac{[\boldsymbol{P}\hat{z}]^T \boldsymbol{S}_\times^T \boldsymbol{t}}{[\boldsymbol{P}\hat{z}]^T \boldsymbol{t}} \\
&= \frac{\hat{z}^T \boldsymbol{S}_\times^T \boldsymbol{t}}{\hat{z}^T \boldsymbol{t}}
\end{aligned}
.
\tag{B6}
$$

Now we use the arctangent addition identity,

$$
\mathrm{atan}u + \mathrm{atan}v = \mathrm{atan}\left(\frac{u+v}{1-uv}\right),
\tag{B7}
$$

where

$$
\frac{u+v}{1-uv} = \frac{\boldsymbol{r}^T \boldsymbol{S}_\times^T \boldsymbol{t}}{\boldsymbol{r}^T \boldsymbol{t}},
\tag{B8}
$$

and

$$
v = \tan\Phi_{\mathrm{tx}} = \frac{\hat{z}^T \boldsymbol{S}_\times^T \boldsymbol{t}}{\hat{z}^T \boldsymbol{t}}.
\tag{B9}
$$

Thus,

$$
\frac{u + \dfrac{\hat{z}^T \boldsymbol{S}_\times^T \boldsymbol{t}}{\hat{z}^T \boldsymbol{t}}}{1 - u\dfrac{\hat{z}^T \boldsymbol{S}_\times^T \boldsymbol{t}}{\hat{z}^T \boldsymbol{t}}} = \frac{\boldsymbol{r}^T \boldsymbol{S}_\times^T \boldsymbol{t}}{\boldsymbol{r}^T \boldsymbol{t}}.
\tag{B10}
$$

Solving for $u = \tan\Phi_{\mathrm{rx}}$, we get,

$$
\tan\Phi_{\mathrm{rx}} = \frac{\dfrac{\boldsymbol{r}^T \boldsymbol{S}_\times^T \boldsymbol{t}}{\boldsymbol{r}^T \boldsymbol{t}} - \dfrac{\hat{z}^T \boldsymbol{S}_\times^T \boldsymbol{t}}{\hat{z}^T \boldsymbol{t}}}{1 + \dfrac{\boldsymbol{r}^T \boldsymbol{S}_\times^T \boldsymbol{t}}{\boldsymbol{r}^T \boldsymbol{t}}\dfrac{\hat{z}^T \boldsymbol{S}_\times^T \boldsymbol{t}}{\hat{z}^T \boldsymbol{t}}}.
\tag{B11}
$$

Simplifying the fraction,

$$
\begin{aligned}
\tan\Phi_{\mathrm{rx}} &= \frac{(\boldsymbol{r}^T \boldsymbol{S}_\times^T \boldsymbol{t})(\hat{z}^T \boldsymbol{t}) - (\hat{z}^T \boldsymbol{S}_\times^T \boldsymbol{t})(\boldsymbol{r}^T \boldsymbol{t})}{(\hat{z}^T \boldsymbol{t})(\boldsymbol{r}^T \boldsymbol{t}) + (\boldsymbol{r}^T \boldsymbol{S}_\times^T \boldsymbol{t})(\hat{z}^T \boldsymbol{S}_\times^T \boldsymbol{t})} \\
&= \frac{\boldsymbol{r}^T \boldsymbol{S}_\times^T \boldsymbol{t}\boldsymbol{t}^T \hat{z} - \boldsymbol{r}^T \boldsymbol{t}\boldsymbol{t}^T \boldsymbol{S}_\times^T \hat{z}}{\boldsymbol{r}^T \boldsymbol{t}\boldsymbol{t}^T \hat{z} - \boldsymbol{r}^T \boldsymbol{S}_\times \boldsymbol{t}\boldsymbol{t}^T \boldsymbol{S}_\times \hat{z}} \\
&= \frac{\boldsymbol{r}^T (\boldsymbol{S}_\times^T \boldsymbol{t}\boldsymbol{t}^T - \boldsymbol{t}\boldsymbol{t}^T \boldsymbol{S}_\times)\hat{z}}{\boldsymbol{r}^T (\boldsymbol{t}\boldsymbol{t}^T - \boldsymbol{S}_\times^T \boldsymbol{t}\boldsymbol{t}^T \boldsymbol{S}_\times)\hat{z}}.
\end{aligned}
\tag{B12}
$$

To further simplify these expressions, we will use identities that arise from the structure of the matrices $\boldsymbol{P}$ and $\boldsymbol{S}_\times$: for any vector $\boldsymbol{v}$,

$$
\begin{aligned}
&\boldsymbol{P}\boldsymbol{v}\boldsymbol{v}^T \boldsymbol{P} - \boldsymbol{S}_\times \boldsymbol{v}\boldsymbol{v}^T \boldsymbol{S}_\times = (\boldsymbol{v}^T \boldsymbol{P}\boldsymbol{v})\boldsymbol{P} \\
&\boldsymbol{P}\boldsymbol{v}\boldsymbol{v}^T \boldsymbol{S}_\times^T + \boldsymbol{S}_\times^T \boldsymbol{v}\boldsymbol{v}^T \boldsymbol{P} = (\boldsymbol{v}^T \boldsymbol{P}\boldsymbol{v})\boldsymbol{S}_\times^T
\end{aligned}
.
\tag{B13}
$$

Because $\boldsymbol{t}$ is defined as orthogonal to $\hat{s}$, $\boldsymbol{t} = \boldsymbol{P}\boldsymbol{t} = \boldsymbol{t}\boldsymbol{P}$, and $\boldsymbol{t}^T \boldsymbol{P}\boldsymbol{t} = \boldsymbol{t}^T \boldsymbol{t}$; therefore,

$$
\begin{aligned}
\tan\Phi_{\mathrm{rx}} &= \frac{\boldsymbol{r}^T (\boldsymbol{S}_\times^T \boldsymbol{t}\boldsymbol{t}^T \boldsymbol{P} - \boldsymbol{P}\boldsymbol{t}\boldsymbol{t}^T \boldsymbol{S}_\times)\hat{z}}{\boldsymbol{r}^T (\boldsymbol{P}\boldsymbol{t}\boldsymbol{t}^T \boldsymbol{P} - \boldsymbol{S}_\times^T \boldsymbol{t}\boldsymbol{t}^T \boldsymbol{S}_\times)\hat{z}} \\
&= \frac{\boldsymbol{r}^T (\boldsymbol{t}^T \boldsymbol{t})\boldsymbol{S}_\times^T \hat{z}}{\boldsymbol{r}^T (\boldsymbol{t}^T \boldsymbol{t})\boldsymbol{P}\hat{z}} = \frac{\boldsymbol{r}^T \boldsymbol{S}_\times^T \hat{z}}{\boldsymbol{r}^T \hat{z}}.\square
\end{aligned}
\tag{B14}
$$

## Appendix C Demonstration of tracking phase wind-up through beam waveguides

To demonstrate the procedure of tracking the differential feed rotation effect through repeated reflections, we define a local coordinate system for azimuth–elevation radio telescopes. The vector $\hat{\boldsymbol{\xi}}$ is the projection of the source unit vector $\hat{s}$ to the local east–north plane of the radio telescope, and the vector $\hat{\boldsymbol{\eta}}$ is the orthogonal complement to $\hat{\boldsymbol{\xi}}$ that forms the right-handed $\hat{\boldsymbol{\xi}}$-$\hat{\boldsymbol{\eta}}$-$\hat{\boldsymbol{u}}$ system. These vectors can be computed as,

$$
\begin{aligned}
\hat{\boldsymbol{\xi}} &= \frac{(\hat{\boldsymbol{u}} \times \hat{s}) \times \hat{\boldsymbol{u}}}{\|\hat{\boldsymbol{u}} \times \hat{s}\|} \\
\hat{\boldsymbol{\eta}} &= \frac{\hat{\boldsymbol{u}} \times \hat{s}}{\|\hat{\boldsymbol{u}} \times \hat{s}\|}
\end{aligned}
.
\tag{C15}
$$

Figure 18 shows $\hat{\boldsymbol{\eta}}$ and $\hat{\boldsymbol{\xi}}$ for a radio telescope with an azimuth–elevation mount. We will use this coordinate system in both the full Nasmyth and beam waveguide examples.

### C.1 The full Nasmyth focus of YEBES40M

We first consider the radio telescope YEBES40M, which regularly participates in VLBI and utilizes a full Nasmyth focus

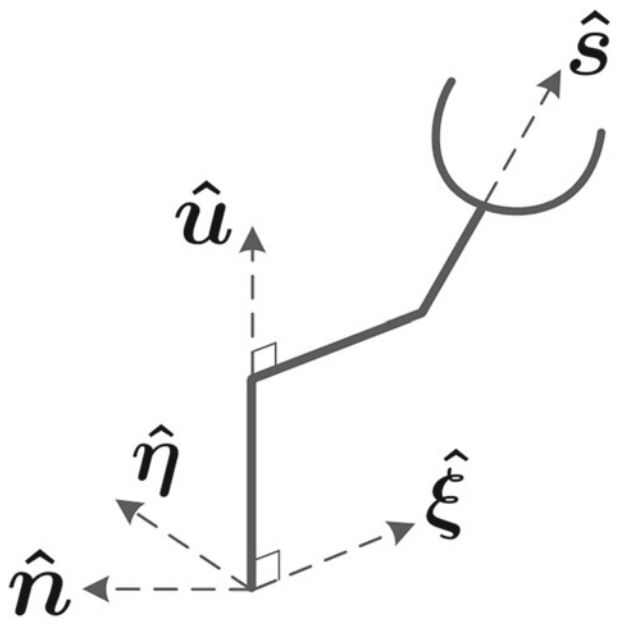

**Fig. 18** A simplified view of the orientation of the $\hat{\xi}$ and $\hat{\eta}$ vectors relative to the local up and north vectors. $\hat{\xi}$ tracks the location of the observed source in the east–north plane, while $\hat{\eta}$ provides the orthogonal pair for the right-handed $\hat{\xi}$-$\hat{\eta}$-$\hat{u}$ system

to allow the telescope to use a variety of receivers for different frequencies. Figure 19 shows the reflections occurring from the primary mirror to the below 20 GHz receivers as detailed in López Fernández et al. (2006) and Tercero et al. (2021). The figure is split into three sections—a back side view of the telescope in the upper left, a side view of the telescope in the upper right, and a view from above the telescope in the bottom. The directions shown in the figure correspond to the $\hat{\xi}$-$\hat{\eta}$-$\hat{u}$ vectors in Fig. 18. From the secondary mirror (M2), a steered planar mirror (M3) reflects incoming light in the negative $\hat{\eta}$ direction to another planar mirror (M4, referred to as M4' in López Fernández et al. (2006) and Tercero et al. (2021)). Note that there is a second configuration in which M2 reflects in the positive $\hat{\eta}$ direction to reach the K, Q, and W band feed horns. This second configuration is not depicted here and is considered out of the scope of the present work. From M4, light is reflected in the negative $\hat{\xi}$ direction to a fifth curved mirror (M5). Mirror M5 reflects back toward mirror M4' but at an angle $\alpha$ in the local down direction. Here the path diverges depending on the observed frequency. An S band feed horn sits on a sliding table with a broadband C-X feed horn, and one of these two feed horns receives light at a time depending on the position of this slide. For either of these two feed horns, the light is received directly from the fifth mirror into the feed horn along the wave vector $\hat{k}_4$. When observing in an S/X band configuration, a dichroic mirror can be placed in the path after mirror M5 to redirect higher-frequency light to a third X band receiver. Lower-frequency radio emission continues to travel to the S band feed horn, while higher-frequency emission is redirected from the planar dichroic mirror M6. From M6, light is reflected back toward mirror M5 but at an angle $\beta$ in the local down direction.

The wave vectors as defined in the figure are given by,

$$
\begin{aligned}
\hat{k}_1 &= -\hat{s} \\
\hat{k}_2 &= -\hat{\eta} \\
\hat{k}_3 &= -\hat{\xi} \\
\hat{k}_4 &= \hat{\xi}\cos\alpha - \hat{u}\sin\alpha \\
\hat{k}_5 &= -\hat{\xi}\cos\beta - \hat{u}\sin\beta.
\end{aligned}
\tag{C16}
$$

The feed horn for an FN focus antenna is in the receiver cabin, which rotates with the telescope in azimuth. The vector primitives representing the aligned and transverse dipoles of the receiver therefore also corotate.

For the S and C-X feed horns, regardless of the angle $\alpha$, the aligned receiver dipole can be defined as $-\hat{\eta}$, while the transverse dipole vector changes orientation based on the angle at which mirror M5 reflects to the feed horn:

$$
\begin{aligned}
\hat{r}^a_{S,C} &= -\hat{\eta} \\
\hat{r}^t_{S,C} &= \hat{u}\cos\alpha + \hat{\xi}\sin\alpha
\end{aligned}
. \tag{C17}
$$

For the X band feed horn receiving emission from the dichroic mirror M6, the aligned receiver dipole is flipped in direction to $\hat{\eta}$. The transverse dipole vector then changes orientation based on the angle mirror M6 makes with the horizontal, $\beta$:

$$
\begin{aligned}
\hat{r}^a_{X} &= \hat{\eta} \\
\hat{r}^t_{X} &= \hat{u}\cos\beta - \hat{\eta}\sin\beta
\end{aligned}
. \tag{C18}
$$

From private communication, the angle $\alpha$ is 51.15 degrees and the angle $\beta$ is 53.87 degrees. In VLBI processing, because the S and C-X band feed horns occur after an odd number of reflections (M5), an additional 180-degree phase offset will be applied by relabeling the detected polarization to the 'on-sky' polarization. In practice, this means that the feed rotation observed at the S and C-X feed horns will be identical to the feed rotation observed at the X feed horn after the dichroic mirror.

Equations 43 and 44 do not straightforwardly produce compact analytical forms for the feed rotation correction, but the full correction is easily implemented in code by applying the equations across the reflections defined by the wave vectors in Equation C16. The reflections from the primary and secondary mirrors do not have to be considered, as M2 reverses the phase flip caused by M1; thus, the original transverse and aligned dipole vectors for the transmitter in Equation 43 can be used at mirror M3.

### C.2 The beam waveguide focus of WARK30M

We also consider the reflections inside the beam waveguide focus of the refurbished telecommunications antenna WARK30M in New Zealand. It is common for telecommunications antennas repurposed as radio telescopes to have this focus type, and WARK30M has a unique slanted beam waveguide (Woodburn et al. 2015). The reflections through the waveguide structure are considered explicitly in Fig. 20, which again shows the view from behind the radio telescope (upper left), to the right side (upper right), and from above the radio telescope (bottom).

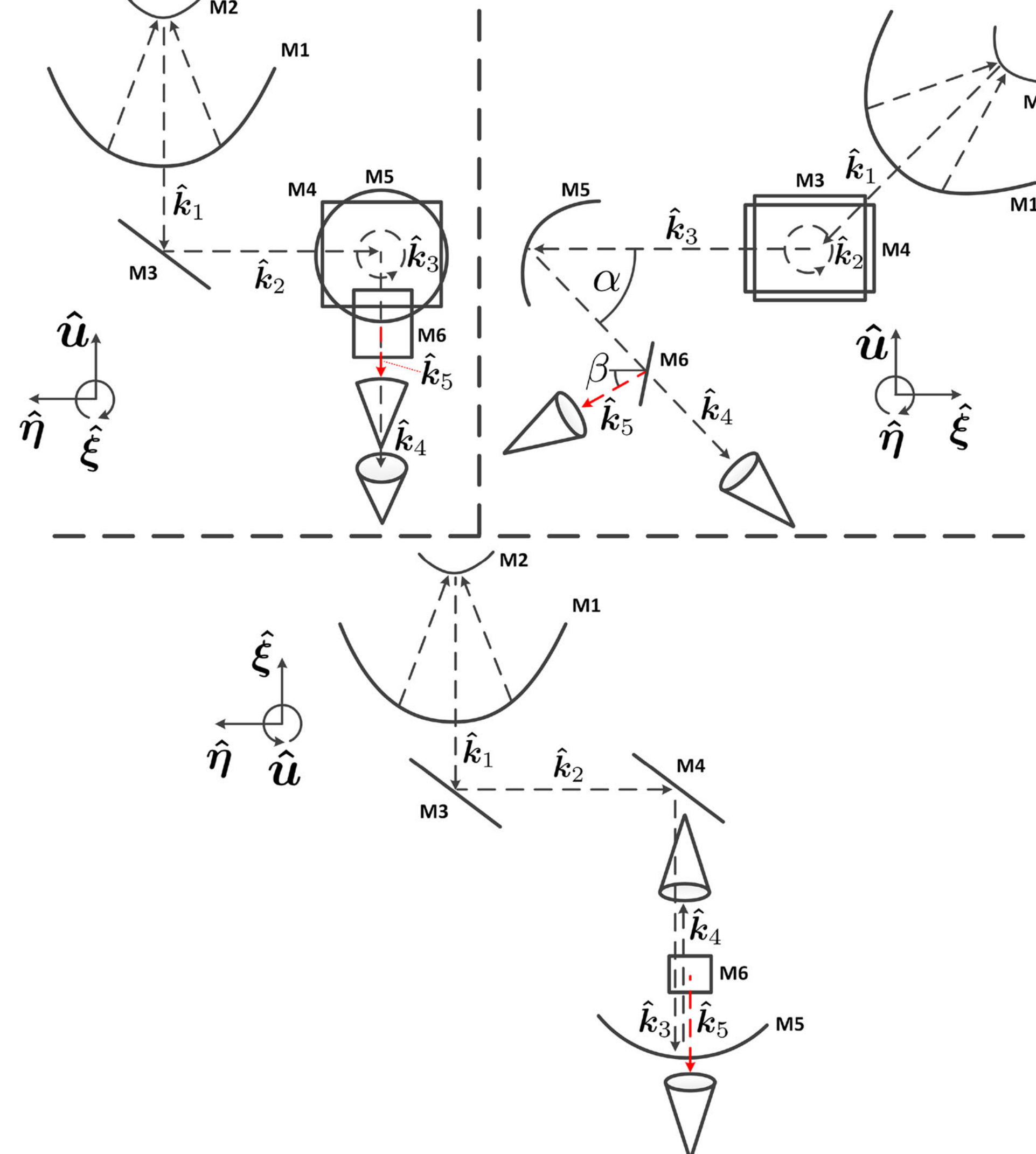

**Fig. 19** A back, side, and top view of the path of a photon entering the YEBES40M receiver cabin. Two reflections occur before light enters the receiver cabin, which rotates in azimuth

The wave vectors depicted in Fig. 20 are given by,

$$
\begin{aligned}
\hat{\boldsymbol{k}}_1 &= -\hat{\boldsymbol{s}} \\
\hat{\boldsymbol{k}}_2 &= -\hat{\boldsymbol{\eta}} \\
\hat{\boldsymbol{k}}_3 &= \frac{-2500\hat{\boldsymbol{\xi}} - 9500\hat{\boldsymbol{u}}}{\|-2500\hat{\boldsymbol{\xi}} - 9500\hat{\boldsymbol{u}}\|}. \\
\hat{\boldsymbol{k}}_4 &= \hat{\boldsymbol{\eta}} \\
\hat{\boldsymbol{k}}_5 &= -\hat{\boldsymbol{u}}
\end{aligned}
\tag{C19}
$$

Note that $\hat{\boldsymbol{k}}_3$ has a component in the negative $\hat{\boldsymbol{\xi}}$ direction and a component in the negative $\hat{\boldsymbol{u}}$ direction. The scaling factors on these components come from length measurements in millimeters from a technical document describing the dimensions of the waveguide shown in Dodson and Rioja (2022). A more typical beam waveguide may have $\hat{\boldsymbol{k}}_3 = -\hat{\boldsymbol{u}}$. Some waveguides have two additional reflections: one from a curved M5 and a final plane mirror directing the light down, M6.

In contrast to the FN mount type, the feed horn for a BWG focus telescope is in a stationary receiver cabin, meaning the feed horn does not rotate in azimuth. The aligned and transverse dipole vectors can simply be taken to as the east and north unit vectors as is done for a GNSS antenna:

$$
\begin{aligned}
\hat{\boldsymbol{r}}^a &= \hat{\boldsymbol{e}} \\
\hat{\boldsymbol{r}}^t &= \hat{\boldsymbol{n}}.
\end{aligned}
\tag{C20}
$$

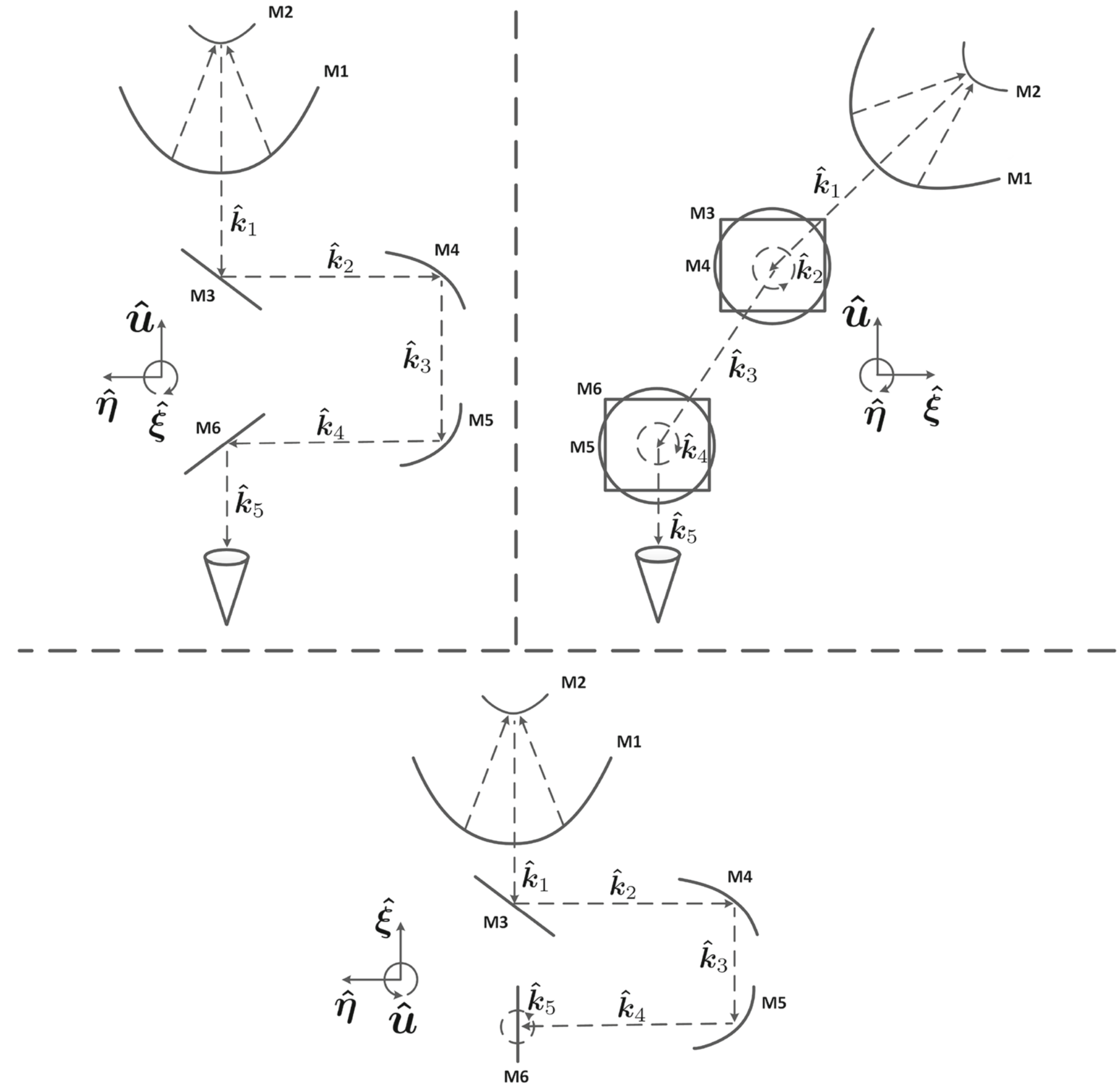

**Fig. 20** A back, side, and top view of the path of a photon entering the WARK30M beam waveguide. Four reflections occur within the beam waveguide before light enters the stationary feed horn at the base of the telescope

## Appendix D A practical guide for differential feed rotation correction

Here we present a self-contained and brief guide to applying the correct model for observations of natural radio sources or satellites with each of the four antenna types of interest: stationary GNSS antennas, radio telescopes with standard focus types, radio telescopes with a full Nasmyth focus, and radio telescopes with a beam waveguide focus.

### D.1 Observations of satellites with a GNSS antenna

**Quantities required for the time epoch of interest:**

- Satellite position, $\boldsymbol{r}_{\text{sat}}$ (Earth-centered, Earth-fixed reference frame)
- Antenna position, $\boldsymbol{r}_{\text{ant}}$ (Earth-centered, Earth-fixed reference frame)
- Satellite orientation quaternion, $q_s$, (if available) OR Sun position, $\boldsymbol{r}_{\text{sun}}$ (Earth-centered, Earth-fixed reference frame)
- North celestial pole, $\hat{z} = [0, 0, 1]^T$ (Earth-centered, Earth-fixed reference frame)

**Steps to compute the differential feed rotation correction:**

1. Find the source unit vector (Equation 2)
2. Compute the projection and cross product operator matrices (Equations 4 and 6)
3. Find the transmitter orientation:
    - If there is an available high-fidelity quaternion model, extract the transmitter aligned and transverse dipole vectors (Equations 24 and 25)
    - Without a quaternion model, instead use the simple nadir-pointing model (Equations 26 and 27). Note that the Sun position must be in the Earth-centered, Earth-fixed reference frame. In many analysis codes, the solar position vector will initially be in a barycentric frame.
4. Compute the transmitter effective dipole vector (Equation 8)

5. Compute the north and east unit vectors representing the aligned and transverse dipole vectors for the GNSS antenna (Equation 28)
6. Find the receiver effective dipole vector (Equation 29)
7. Compute the differential feed rotation (Equation 7)
8. If this is not the first observation of the transmitter, adjust the differential feed rotation to the cycle of the previous correction (Equation 9)
9. Subtract the differential feed rotation from the phase measurement for this epoch

## D.2 Observations of satellites with a radio telescope

**Quantities required for the time epoch of interest:**

- Satellite position, $\boldsymbol{r}_{\text{sat}}$ (Earth-centered, Earth-fixed reference frame)
- Antenna position, $\boldsymbol{r}_{\text{ant}}$ (Earth-centered, Earth-fixed reference frame)
- Telescope fixed-axis vector, $\hat{\boldsymbol{a}}_{\text{ant}}$ (Earth-centered, Earth-fixed reference frame)
- Satellite orientation quaternion, $q_s$, (if available) OR Sun position, $\boldsymbol{r}_{\text{Sun}}$ (Earth-centered, Earth-fixed reference frame)
- North celestial pole, $\hat{z} = [0\ 0\ 1]^T$ (Earth-centered, Earth-fixed reference frame)

**Steps to compute the differential feed rotation correction:**

1. Find the source unit vector (Equation 2)
2. Compute the projection and cross product operator matrices (Equations 4 and 6)
3. Find the transmitter orientation:
    - If there is an available high-fidelity quaternion model, extract the transmitter aligned and transverse dipole vectors (Equations 24 and 25)
    - Without a quaternion model, instead use the simple nadir-pointing model (Equations 26 and 27). Note that the Sun position must be in the Earth-centered, Earth-fixed reference frame. In many analysis codes, the solar position vector will initially be in a barycentric frame.
4. Compute the transmitter effective dipole vector (Equation 8)
5. Find the receiver effective dipole vector from the telescope fixed-axis vector (Equation 10)
6. Compute the differential feed rotation (Equation 7)
7. If the telescope has an FN focus, add or subtract the elevation angle depending on the handedness of the third reflection (Equation 38)
8. If the telescope has a BWG focus, add or subtract the elevation angle and subtract or add the azimuth angle depending on the handedness of the third reflection (Equation 39)
9. If this is not the first observation of the transmitter, adjust the differential feed rotation to the cycle of the previous correction (Equation 9)
10. Subtract the differential feed rotation from the phase measurement for this epoch (RHCP) or add the differential feed rotation to the phase measurement (LHCP)

## D.3 Observations of natural radio sources with a GNSS antenna

**Quantities required for the time epoch of interest:**

- Source unit vector, $\hat{\boldsymbol{s}}$ (barycentric inertial reference frame)
- Rotation matrix, $\boldsymbol{R}$, from the barycentric inertial reference frame (i.e., International Celestial Reference System) to the Earth-centered, Earth-fixed reference frame (i.e., International Terrestial Reference System)
- Antenna position, $\boldsymbol{r}_{\text{ant}}$ (Earth-centered, Earth-fixed reference frame)
- North celestial pole, $\hat{z} = [0\ 0\ 1]^T$ (Earth-centered, Earth-fixed reference frame)

**Steps to compute the differential feed rotation correction:**

1. Either transform the source unit vector to the Earth-fixed terrestrial reference frame, $\hat{\boldsymbol{s}}^{\text{TRF}} = \boldsymbol{R}\hat{\boldsymbol{s}}^{\text{CRF}} = \boldsymbol{R}[\cos\alpha\cos\delta\ \sin\alpha\cos\delta\ \sin\delta]^T$ for right ascension $\alpha$ and declination $\delta$, or transform the antenna position and north celestial pole to the celestial reference frame: $\boldsymbol{r}_{\text{ant}}^{\text{CRF}} = \boldsymbol{R}^T\boldsymbol{r}_{\text{ant}}$, $\hat{z}^{\text{CRF}} = \boldsymbol{R}^T\hat{z}$
2. Compute the projection and cross product operator matrices (Equations 4 and 6)
3. Compute the transmitter effective dipole vector (Equation 11)
4. Compute the north and east unit vectors representing the aligned and transverse dipole vectors for the GNSS antenna (Equation 28)
5. Find the receiver effective dipole vector (Equation 29)
6. Compute the differential feed rotation (Equation 7)
7. If this is not the first observation of the transmitter, adjust the differential feed rotation to the cycle of the previous correction (Equation 9)
8. Subtract the differential feed rotation from the phase measurement for this epoch

## D.4 Observations of natural radio sources with a radio telescope

**Quantities required for the time epoch of interest:**

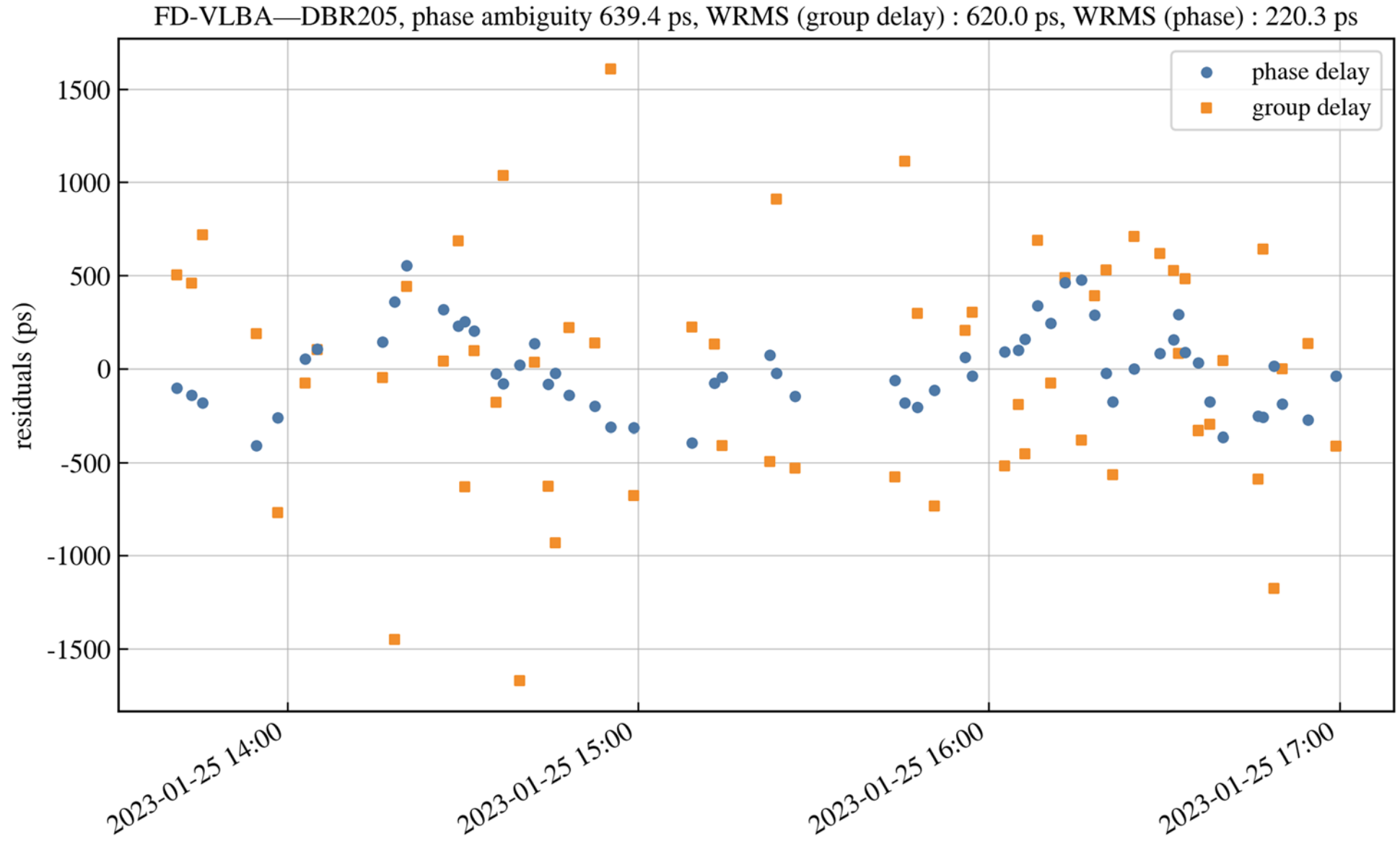

**Fig. 21** Postfit residuals for the group delay-only positioning solution and a phase delay-only solutions without differential feed rotation correction

**Table 6** Phase delay baseline vectors with and without differential feed rotation with components and associated uncertainties in millimeters for the baseline DBR205–FD-VLBA

| Observable | $E$ | $\sigma_E$ | $N$ | $\sigma_N$ | $U$ | $\sigma_U$ | $L$ | $\sigma_L$ |
|---|---|---|---|---|---|---|---|---|
| Phase delay (corrected feed rotation) | 39447.4 | 1.3 | 60630.2 | 1.0 | 12996.4 | 2.7 | 73491.7 | 1.1 |
| Phase delay (uncorrected feed rotation) | 39508.6 | 21.0 | 60467.5 | 15.4 | 12789.0 | 41.2 | 73354.0 | 17.4 |

- Source unit vector, $\hat{\boldsymbol{s}}$ (barycentric inertial reference frame)
- Rotation matrix, $\boldsymbol{R}$, from barycentric inertial reference frame (i.e., International Celestial Reference System) to Earth-centered, Earth-fixed reference frame (i.e., International Terrestial Reference System)
- Telescope fixed-axis vector, $\hat{\boldsymbol{a}}_{\text{ant}}$ (Earth-centered, Earth-fixed reference frame)
- North celestial pole, $\hat{z} = [0, 0, 1]^T$ (Earth-centered, Earth-fixed reference frame)

**Steps to compute the differential feed rotation correction:**

1. Either transform the source unit vector to the Earth-fixed terrestrial reference frame, $\hat{\boldsymbol{s}}^{\text{TRF}} = \boldsymbol{R}\hat{\boldsymbol{s}}^{\text{CRF}} = \boldsymbol{R}[\cos\alpha\cos\delta\ \sin\alpha\cos\delta\ \sin\delta]^T$ for right ascension $\alpha$ and declination $\delta$, or transform the fixed-axis vector and north celestial pole to the celestial reference frame: $\hat{\boldsymbol{a}}^{\text{CRF}} = \boldsymbol{R}^T\hat{\boldsymbol{a}}, \hat{z}^{\text{CRF}} = \boldsymbol{R}^T\hat{z}$
2. Compute the projection and cross product operator matrices (Equations 4 and 6)
3. Compute the transmitter effective dipole vector (Equation 11)
4. Find the receiver effective dipole vector from the telescope fixed-axis vector (Equation 10)
5. Compute the differential feed rotation (Equation 7)
6. If the telescope has an FN focus, add or subtract the elevation angle depending on the handedness of the third reflection (Equation 38)
7. If the telescope has a BWG focus, add or subtract the elevation angle and subtract or add the azimuth angle depending on the handedness of the third reflection (Equation 39)
8. If this is not the first observation of the transmitter, adjust the differential feed rotation to the cycle of the previous correction (Equation 9)
9. Subtract the differential feed rotation from the phase measurement for this epoch (RHCP) or add the differential feed rotation to the phase measurement (LHCP)

## Appendix E Results of vector baseline estimation with the differential feed rotation correction omitted

Figure 13 shows the postfit residuals for the baseline FD-VLBA–DBR205 from the group delay-only positioning solution shown in Section 4.1 with a phase delay-only positioning solution in which the measurements have not been corrected for the differential feed rotation effect. We used the same procedure to resolve the integer ambiguities, producing a phase delay solution with as low WRMS as possible. It is easy to see that the solution is much poorer without this correction, as the WRMS of the phase delay residuals after the least squares adjustment is 220.3 ps, about 16 times larger than the WRMS of the phase delays when the differential feed rotation is corrected (Fig. 21).

Table 6 compares the east–north–up baseline components and uncertainties of the phase delay-only baseline vector with and without the differential feed rotation correction. The two baseline vectors disagree at the level of tens of centimeters—far larger than the formal uncertainties. This shows the importance of the correction to obtaining a good baseline vector estimate.

**Acknowledgements** The authors sincerely thank José Antonio López-Pérez, Javier González, and Félix Tercero-Martínez of the Yebes Observatory for helpful conversation and details of the Yebes 40-meter radio telescope and its receiver cabin setup. The authors also thank David Munton for many helpful comments on the content of the manuscript, Walter Brisken at NRAO for help in coordinating the VLBA observation, Sharyl Byram for sponsoring USNO observing time, and the Fort Davis NRAO technicians Julian Wheat and Juan De Guia as well as the McDonald observatory technicians Eusebio 'Chevo' Terrazas and Renny Spencer who provided operational support. The NRAO is a facility of the National Science Foundation operated under cooperative agreement by Associated Universities, Inc. The authors acknowledge use of the VLBA under the USNO's time allocation. This work made use of the Swinburne University of Technology software correlator, developed as part of the Australian Major National Research Facilities Programme and operated under license.

**Author Contributions** JS adapted the feed rotation model and wrote the manuscript. JS, JY, LP, KH, and RJ collected the satellite observations. JS and LP processed the experimental data. JY, LP, and SB supervised the research. All authors read and approved the final manuscript.

**Funding** We gratefully acknowledge the funding provided for this work under NASA Grant 80NSSC24K0828.
ARL:UT Research and Development supported work not covered under the NASA grant.

**Data Availability** All data including the FITS data for the DBR205–FD-VLBA experiment and the fringe fitting results for the experiments EB050 and GM074Z, as well as the custom code used to produce the results in this manuscript are available at the Texas Data Repository Skeens et al. (2025). Source code including the fringe fitting software PIMA and high-fidelity time delay software VTD are available at https://github.com/nasa/sgdass with further documentation at https://astrogeo.smce.nasa.gov/sgdass/. A routine for differential feed rotation correction can be found in the function VTD_CALC_PHASE_WINDUP at https://github.com/nasa/sgdass/blob/main/vtd/src/vtd_calc_pco.f.
FITS data for the WARK30M–WARK12M experiment, and GPS final orbit products including ephemerides and clock solutions were downloaded from the Crustal Dynamics Data Information System (Noll 2010; NASA Crustal Dynamics Data Information System 1992). The software correlator DiFX is available at https://github.com/difx/difx.

## Declarations

**Conflict of interest** The authors declare that they have no Conflict of interest.

## References

Beyerle G (2009) Carrier phase wind-up in gps reflectometry. GPS Solut 13(3):191–198. https://doi.org/10.1007/s10291-008-0112-1

Bondi M, Pérez-Torres M, Piconcelli E et al (2016) Unveiling the radio counterparts of two binary agn candidates: J1108+ 0659 and j1131–0204. Astron Astrophys 588:A102

Cappallo R (2014) Correlating and fringe-fitting broadband vgos data. In: International VLBI Service for Geodesy and Astrometry 2014 General Meeting Proceedings: VGOS: The New VLBI Network, pp 91–96

Cotton W (1993) Calibration and imaging of polarization sensitive very long baseline interferometer observations. Astron J 106(3):1241–1248. https://doi.org/10.1086/116723

Deller A, Tingay SJ, Bailes M et al (2007) DiFX: A software correlator for very long baseline interferometry using multiprocessor computing environments. Publ Astron Soc Pac 119(853):318–336. https://doi.org/10.1086/513572

Deller AT, Brisken WF, Phillips CJ et al (2011) Difx-2: a more flexible, efficient, robust, and powerful software correlator. Publ Astron Soc Pac 123(901):275. https://doi.org/10.1086/658907

Delva P, Altamimi Z, Blazquez A et al (2023) Genesis: co-location of geodetic techniques in space. Earth Planets Space 75(1):5. https://doi.org/10.1186/s40623-022-01752-w

Dodson R (2009) On the solution of the polarisation gain terms for vlbi data collected with antennas having nasmyth or ew mounts. arXiv preprint arXiv:0910.1707

Dodson R, Rioja MJ (2022) Feed rotation corrections for antennas having beam waveguide mounts. arXiv preprint arXiv:2210.13381

Greisen EW (2003) AIPS, the VLA, and the VLBA. In: Heck A (ed) Information Handling in Astronomy - Historical Vistas, p 109, https://doi.org/10.1007/0-306-48080-8_7

van Haarlem MP, Wise MW, Gunst A et al (2013) Lofar: The low-frequency array. Astron Astrophys 556:A2

Jackson JD (2021) Classical electrodynamics. John Wiley & Sons

Jaron F, Nothnagel A (2019) Modeling the vlbi delay for earth satellites. J Geodesy 93(7):953–961. https://doi.org/10.1007/s00190-018-1217-0

Skeens J, Petrov L (2024) Implementing a VLBI Time Delay Model for Earth-orbiting Satellites: Partial derivatives and Verification. Tech. rep., Goddard Space Flight Center, https://ntrs.nasa.gov/citations/20240007790, nASA Technical Reports Server Document 20240007790

Karatekin O, Sert H, Dehant V, et al (2023) Vlbi signals transmitted from earth orbiting satellites. In: 26th European VLBI Group for Geodesy and Astronomy Working Meeting

López Fernández J, Gómez González J, Barcía Cáncio A (2006) Radio telescope engineering: the 40m ign antenna. Lecture Notes Essays Astrophys 2:257–270

Loyer S, Banville S, Geng J et al (2021) Exchanging satellite attitude quaternions for improved gnss data processing consistency. Adv Space Res 68(6):2441–2452. https://doi.org/10.1016/j.asr.2021.04.049

Marti-Vidal I, Roy A, Conway J et al (2016) Calibration of mixed-polarization interferometric observations-tools for the reduction of interferometric data from elements with linear and circular polarization receivers. Astron Astrophys 587:A143

Marti-Vidal I, Mus A, Janssen M et al (2021) Polarization calibration techniques for the new-generation vlbi. Astron Astrophys 646:A52

McCallum L, Schunck D, McCallum J, et al (2024) An instrument to link global positioning to the universe–observing gnss satellites with the australian vlbi array. arXiv preprint arXiv:2412.07020

Montenbruck O, Schmid R, Mercier F et al (2015) Gnss satellite geometry and attitude models. Adv Space Res 56(6):1015–1029

NASA Crustal Dynamics Data Information System (1992) GNSS Final Combined Orbit Solution Product [Dataset]. https://doi.org/10.5067/GNSS/GNSS_IGSORB_001

Noll CE (2010) The crustal dynamics data information system: A resource to support scientific analysis using space geodesy. Adv Space Res 45(12):1421–1440. https://doi.org/10.1016/j.asr.2010.01.018

Nothnagel A (2024) Elements of Geodetic and Astrometric Very Long Baseline Interferometry. TU Wien, Vienna, https://www.vlbi.at/data/publications/Nothnagel_Elements_of_VLBI.pdf

Petrov L, Kovalev Y, Fomalont E et al (2011) The Very Long Baseline Array Galactic Plane Survey -VGaPS. Ast J 142:35. https://doi.org/10.1088/0004-6256/142/2/35

Petrov L, Natusch T, Weston S et al (2015) First scientific vlbi observations using new zealand 30 meter radio telescope wark30m. Publ Astron Soc Pac 127(952):516

Petrov L, York J, Skeens J, et al (2024) A concept of precise vlbi/gnss ties with micro-vlbi. In: Freymueller JT, Sánchez L (eds) Gravity, Positioning and Reference Frames. Springer Nature Switzerland, Cham, pp 147–152, https://doi.org/10.1007/1345_2023_211

Rhodes B (2023) sgp4: Satellite orbit prediction using sgp4. https://pypi.org/project/sgp4/, version 2.21

Rife JH, Weaver BM, Bogner T et al (2021) Combined effects of geometric rotation and antenna calibration patterns on gnss phase windup. IEEE Trans Aerosp Electron Syst 57(5):3185–3197

Schunck D, McCallum L, Molera Calvés G (2024) On the integration of vlbi observations to genesis into global vgos operations. Remote Sens. https://doi.org/10.3390/rs16173234

Skeens J, York J, Petrov L et al (2023) First observations with a gnss antenna to radio telescope interferometer. Radio Sci 58(8):e2023RS007734. https://doi.org/10.1029/2023RS007734

Skeens J, York J, Petrov L, et al (2024) Extracting geodetic data from gnss-vlbi co-observation. arXiv preprint arXiv:2410.14834

Skeens J, York J, Petrov L, et al (2025) A unified model of feed rotation in radio telescopes and GNSS antennas [Dataset]. https://doi.org/10.18738/T8/TPBQP7

Tercero F, López-Pérez J, Gallego J et al (2021) Yebes 40 m radio telescope and the broad band nanocosmos receivers at 7 mm and 3 mm for line surveys. Astron Astrophys 645:A37

Thompson AR, Moran JM, Swenson GW, Jr. (2017) Interferometry and Synthesis in Radio Astronomy, 3rd Edition. Springer, https://doi.org/10.1007/978-3-319-44431-4

Thompson JM, Moran AR, Swenson GW Jr. (2001) Interferometry and Synthesis in Radio Astronomy, 2nd edn. John Wiley and Sons, New York

Vallado D, Crawford P (2008) Sgp4 orbit determination. In: AIAA/AAS Astrodynamics Specialist Conference and Exhibit, p 6770

Woodburn L, Natusch T, Weston S et al (2015) Conversion of a new zealand 30-metre telecommunications antenna into a radio telescope. Publ Astron Soc Austral 32:e017

Wu J, Wu S, Hajj G et al (1993) Effects of antenna orientation on gps carrier phase. Manuscr Geod 18:91–98